\documentclass[10pt,aps,prmaterials,amsfonts,amsmath,amssymb,raggedbottom,longbibliography,reprint,citeautoscript,bibnotes]{revtex4-2}
\usepackage[usenames,dvipsnames]{color}
\usepackage{graphicx,microtype}
\usepackage[version=4]{mhchem} 
\usepackage{booktabs} 
\usepackage{multirow} 
\usepackage{enumitem}
\usepackage{lmodern}
\usepackage{placeins} 
\usepackage[bookmarks=false,colorlinks]{hyperref}
\hypersetup{linkcolor=magenta,citecolor=MidnightBlue,filecolor=Plum,urlcolor=MidnightBlue}
\usepackage[all]{hypcap} 
\usepackage{soul}

\newcommand\blank{\rule[-.2ex]{0.2cm}{.4pt}}

\newcommand{\supf}{\textcolor{DarkOrchid}{Fig.}\,S}


\begin{document}
\title{Decoratypes: An Extensible Crystal Taxonomy for Machine Learning-Guided Materials Discovery} 

\author{Kyle D.\ Miller}
\affiliation{Department\;of\;Materials\;Science\;and\;Engineering, Northwestern University, Evanston, Illinois  60208, USA}

\author{Michele Campbell}
\affiliation{Department\;of\;Materials\;Science\;and\;Engineering, Northwestern University, Evanston, Illinois  60208, USA}

\author{Danilo Puggioni}
\email{danilo.puggioni@northwestern.edu}
\affiliation{Department\;of\;Materials\;Science\;and\;Engineering, Northwestern University, Evanston, Illinois  60208, USA}

\author{James M.\ Rondinelli}
\email{jrondinelli@northwestern.edu}
\affiliation{Department\;of\;Materials\;Science\;and\;Engineering, Northwestern University, Evanston, Illinois  60208, USA}

\begin{abstract}
We introduce \textit{decoratypes} as a structure taxonomy that classifies compounds based on site decorations of specific structural prototypes. Building on this foundation, a ferroelectric materials discovery framework is developed, integrating decoratypes with an active learning approach to accelerate exploration. In addition, six novel ferroelectric candidates are predicted, including three strain-activated ferroelectrics and three strain-activated hyperferroelectrics.
These findings highlight the potential of the decoratype taxonomy to enhance our understanding of structure-driven material properties and facilitate the discovery of promising yet underexplored regions of chemical space.
\end{abstract}

\date{\today}
\maketitle

\section{Introduction}

The quest to discover new materials has fueled crystal structure classification efforts for centuries. 
By grouping structures with similar atomic arrangements, researchers can more easily identify trends and structure-property relationships, accelerating the discovery of functional materials. 
As both experimental and computational materials discovery accelerate, the demand for efficient sorting of new materials becomes increasingly critical. 
Recent studies have already generated an unprecedented number of theoretical materials \cite{szymanski2023, merchant2023} and still others are exploring the application of generative machine learning approaches to materials discovery \cite{alverson2024, luu2023}, promising to further expand the materials discovery pipeline.
A comprehensive classification scheme is needed to help determine novelty and utility \cite{Cheetham2024}, and contextualize these theoretical materials within existing materials families.

Several notable efforts have been made to develop crystal structure classification schemes, providing a framework for processing and archiving data generated from X-ray crystallography.
One of the earliest attempts at systematic crystal structure classification was the Strukturbericht, developed by Ewald, Hermann, and Straumanis in the 1930s \cite{hermann1931}. This scheme assigns a letter-number combination to each structure type, with the letter indicating the crystal system and the number serving as an arbitrary identifier. 
Although the Strukturbericht laid important groundwork, its lack of a hierarchical organization limits its utility for identifying structural similarities across broad chemical spaces.
In the 1960s, Pearson introduced a more extensible notation that combined the crystal system, centering, and number of atoms in the unit cell \cite{pearson1958}. 
Pearson's notation enables a more granular classification of crystal structures and 
forms the basis for Pearson's Crystal Data database \cite{villars2008}, an expansion of the Pauling File project \cite{villars2004}.
In the 1990s, the International Union of Crystallography proposed a standard nomenclature \cite{lima-de-faria1990}, encouraging the development of increasingly sophisticated crystal structure taxonomies. 

The development of digital databases has greatly expanded the scope of crystal structure classification, cataloging hundreds of thousands of inorganic compounds.
The Inorganic Crystal Structure Database (ICSD), released in 1978, aims to catalog all published inorganic crystal structures \cite{bergerhoff1987, zagorac2019}. 
It introduced structure types in 2007 \cite{allmann2007} using a hierarchical classification scheme \cite{lima-de-faria1990}. It first assigns an isopointal structure type, which refers to structures that share the same space group and Wyckoff sites. Specifically, they have identical symmetry and atomic site configurations but may differ in composition. 
It then subdivides these into isoconfigurational structure types, which are isopointal but also exhibit similar atomic arrangements and bonding characteristics. 
This classification is further refined based on structural descriptors, including unit cell dimensions and crystal chemistry.

More recently, the Automatic Flow for Materials Discovery (AFLOW) crystal prototype encyclopedia has emerged as a powerful tool for surveying the crystal structure space \cite{mehl2017}. A prototype, in this context, refers to a standard or model crystal structure that serves as a representative example of a specific structural type, which can be used to classify other crystal structures. By using a combination of symmetries and local atomic environment descriptors, the AFLOW prototype scheme classifies structures with similar prototypes, regardless of their chemical composition or atomic coordinates. This abstract representation of crystal structures has already succeeded in enabling machine learning-based predictions of materials properties \cite{oses2020, gubaev2019, ziletti2018}.

Prototyping systems provide a standardized framework for organizing and interpreting structural data, enabling the use of informatics and machine learning to accelerate materials discovery. 
By systematically describing the atomic structure of interrelated compounds, researchers can identify trends that lead to the development of design rules for engineering new materials. 
Since many key material properties are governed by atomic-scale structure, the classification of crystal structures serves as a powerful tool in materials science. However, such classifications are most informative when coupled with information about the occupancy of atomic species within the unit cell. 

Here, we propose a unified framework, dubbed \textit{decoratypes} -- a portmanteau of decoration and prototype, to describe the set of all possible decorations of an isopointal structure type. 
It integrates both structural and chemical properties, allowing for a more comprehensive approach to understanding and designing crystalline materials.
Within this framework, we introduce a subset classification termed \textit{polaritype}, which generalizes anti-structures (or inverse structures) by categorizing crystal structures based on mappings between ionic polarity and Wyckoff site occupancy
Specifically, we target an anti-structure family containing known ferroelectric and hyperferroelectric materials: anti-Ruddlesden-Popper structures \cite{markov2021, kang2020}. 
We then develop a machine learning workflow that combines polaritype-based screening for computational discovery of ferroelectric and hyperferroelectric compounds in this family.

\section{The Decoratype Taxonomy}

\begin{figure}
    \centering
    \includegraphics[width=0.9\linewidth]{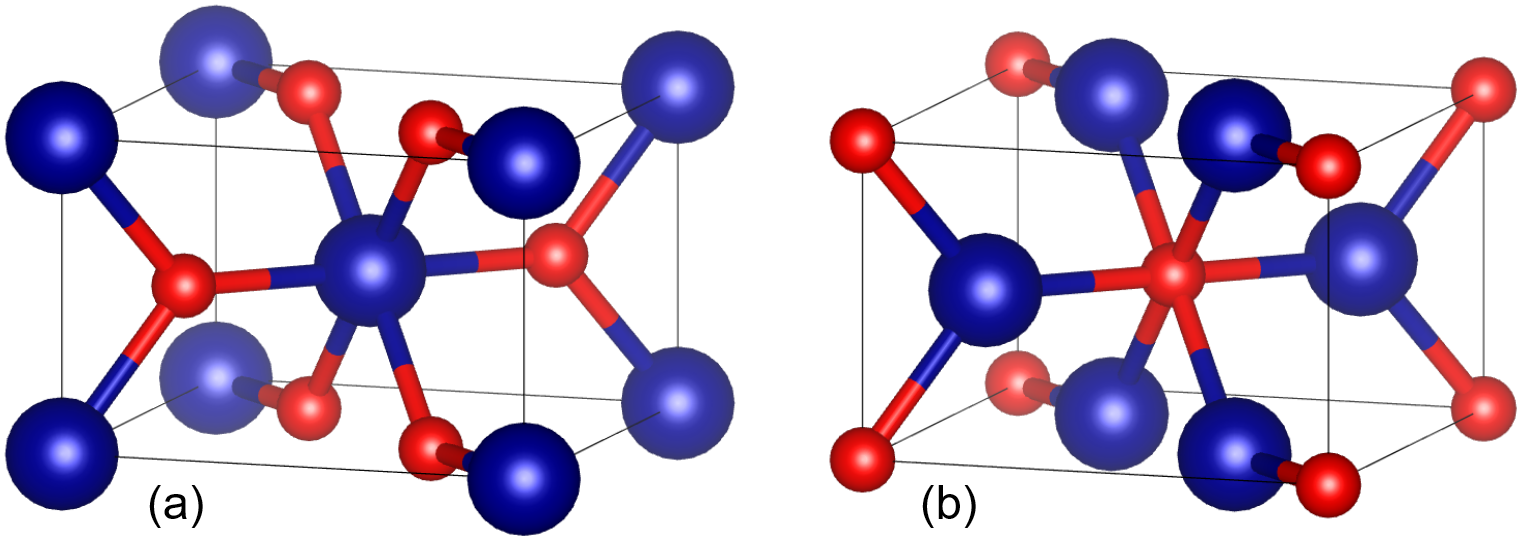}
    \caption{Example of the binary polaritype family within the rutile structure. Unit cells of (a) rutile (A$^+$B$^-_2$ or \text{A2B\blank{}tP6\blank{}136\blank{}f\blank{}a\blank{}\blank{}-\blank{}+}) and (b) anti-rutile structures (A$^-$B$^+_2$ or \text{A2B\blank{}tP6\blank{}136\blank{}f\blank{}a\blank{}\blank{}+\blank{}-}). Large, blue spheres indicate cations and small, red spheres indicate anions.}
    \label{fig:deco-antirutile}
\end{figure}

To introduce the decoratype taxonomy, we first define anti-structures, then describe how the decoratype framework arises as a natural generalization of the anti-structure classification scheme.
An anti-structure, also called an inverse structure, is broadly  a crystal that is isopointal to another crystal and has at least one site occupied by a cation rather than an anion and one site occupied by an anion rather than a cation relative to the first crystal. 
For example, anti-rutile Ti$_2$N is the anti-structure to rutile TiO$_2$ (space group 136, $P4_2/mnm$), because the anion and cation positions (Wyckoff sites $2a$ and $4f$) have been exchanged, as shown in \autoref{fig:deco-antirutile}. 
Evidence suggests that the materials subspace of anti-structures is rich in non-oxide and heteroanionic functional materials such as the electride Ca$_2$N \cite{tang2018}, the battery materials Li$_4$O(Cl$_{1-b}$Br$_b$)$_2$ \cite{jalem2021}, Li$_2$Fe$_2$S$_2$O, and Li$_4$Fe$_3$S$_3$O \cite{zhu2022}, the electrocatalysts Co$_2$P \cite{callejas2015} and Co$_2$N \cite{sun2020}, the MXene Ti$_2$C \cite{soundiraraju2017} and the metal-insulator transition compound  Cu$_3$N \cite{yu2005,yu2005a,liu-xiang2006,wang2006,wosylus2009,zhao2010,ghoohestani2014,kuzmin2018}. 

\begin{figure}
    \centering
    \includegraphics[width=1\linewidth]{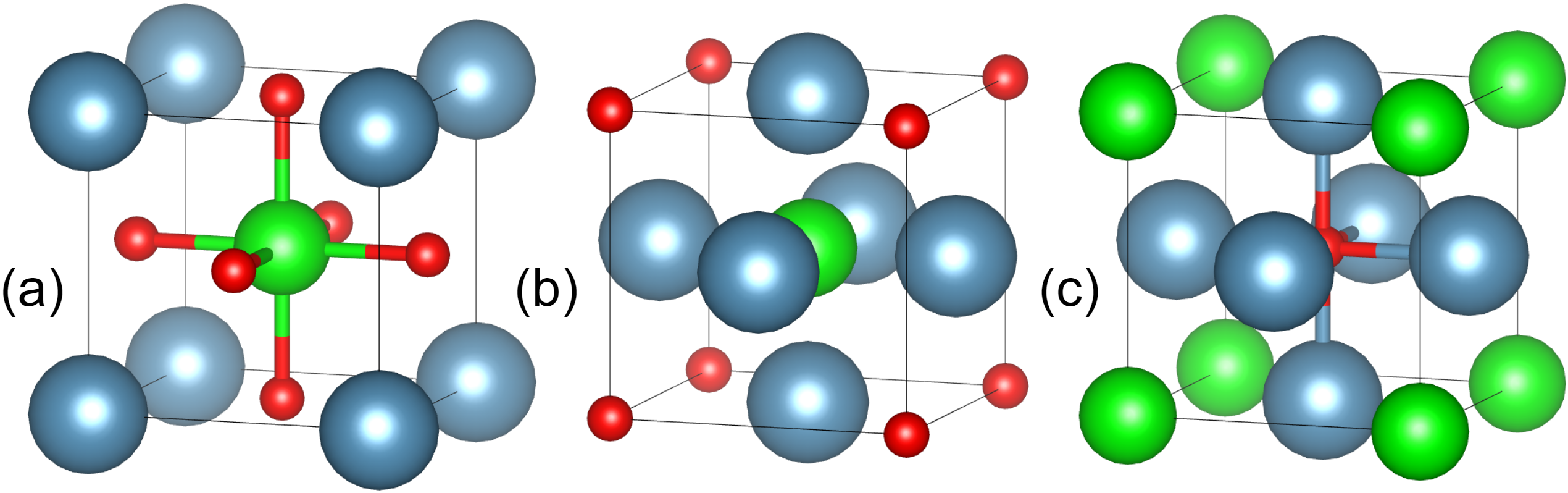}
    \caption{Example of a partial ternary polaritype family within the cubic perovskite structural prototype with theoretical member formulae: (a) A$^+$B$^+$C$^-_3$ (\text{ABC3\blank{}cP5\blank{}221\blank{}a\blank{}b\blank{}c\blank{}\blank{}+\blank{}+\blank{}-}), (b) A$^+$B$^-$C$^+_3$ (\text{ABC3\blank{}cP5\blank{}221\blank{}a\blank{}b\blank{}c\blank{}\blank{}+\blank{}-\blank{}+}), and (c) A$^-$B$^+$C$^+_3$ (\text{ABC3\blank{}cP5\blank{}221\blank{}a\blank{}b\blank{}c\blank{}\blank{}-\blank{}+\blank{}+}). Large blue and green spheres indicate cations and small red spheres indicate anions.}
    \label{fig:deco-perovskite}
\end{figure}

Although there are many examples of anti-structures, these studies are typically limited in scope to a single structural family \cite{callejas2015, gao2021, jalem2021, markov2021, walters2024}.
To date, no standardized database or comprehensive nomenclature has been established to systematically describe anti-structure-like relationships.
%
%
While useful for certain crystal types, the binary classification of structures and anti-structures is inherently limited, as it is well-defined only for ionic compounds featuring exactly two distinct occupied Wyckoff sites.
Binary compounds like rutile,  \autoref{fig:deco-antirutile}, can be mapped to a binary term.
Ternary compounds, however, often contain more than two possible ``anti-structures'' as shown in \autoref{fig:deco-perovskite}.
For an arbitrary ionic compound with $w$ Wyckoff sites, the definition of anti-structure is insufficient to describe the $2^{w} - 2$ possible anion/cation decorations of $w$ sites (excluding all-anion and all-cation decorations). 

\begin{figure}[b]
    \centering
    \includegraphics[width=0.98\linewidth]{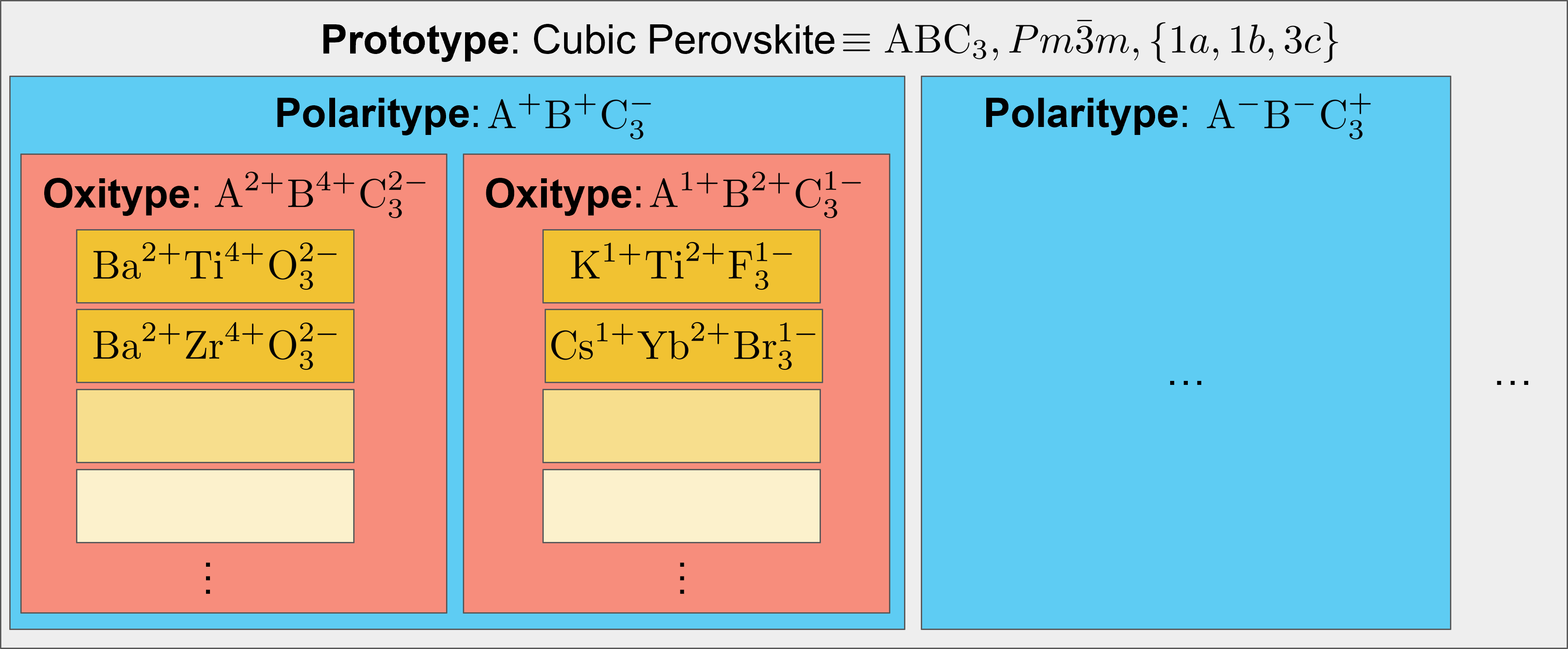}
    \caption{Example of charge-based decoratype hierarchy, i.e., polaritype (blue) $\supset$ oxitype (red) $\supset$ compounds (yellow), for cubic perovskites. 
    }
    \label{fig:schematic-chg-decoratypes}
\end{figure}

\subsection{Definition}
To remove this ambiguity, we propose \textbf{decoratypes} (\textbf{decora}tion + proto\textbf{types}) as an extensible taxonomy to comprehensively describe all possible decorations of a structural prototype with arbitrary site-based properties such as polarity, oxidation state, magnetic moment, and other site-based properties. 
\autoref{fig:schematic-chg-decoratypes} displays this hierarchical classification scheme using two charge-related decoratypes and a nomenclature we formally define below. Here, the structural prototype is cubic perovskite. It is divided into \emph{polaritype} classes, representing groups of cubic perovskite compounds with specific cation/anion decorations of the lattice. \emph{Oxitype} classes further subdivide each polaritype into groups of compounds with specific oxidation-state decorations of the lattice.

\begin{figure*}
    \centering
    \includegraphics[width=0.67\linewidth]{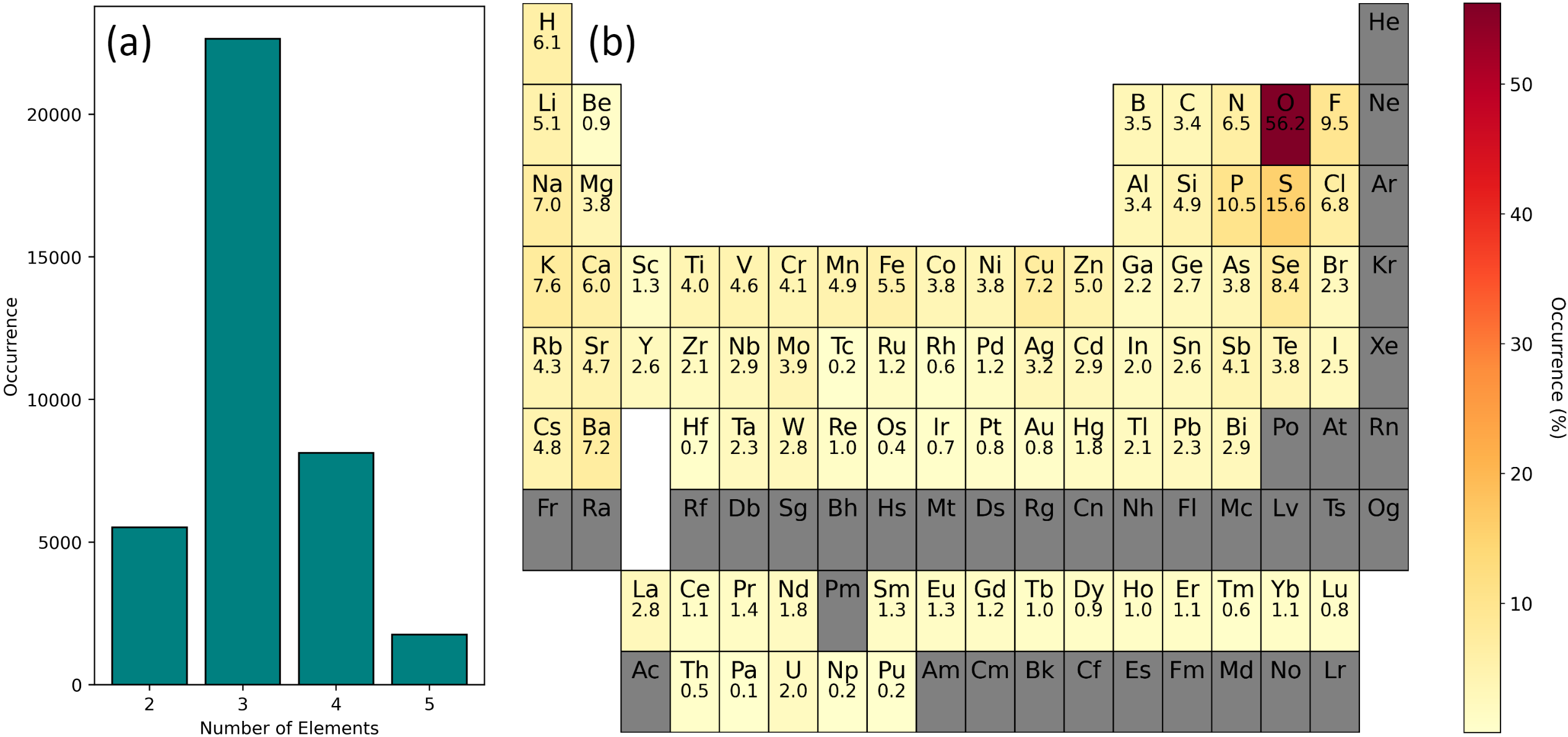}
    \caption{(a) Histogram showing the population by number of elements and (b) a heatmap showing the frequency of elements in the MP data set.}
    \label{eda}
\end{figure*}

\subsection{Nomenclature}

\begin{itemize}[label={}, leftmargin=0cm]
    \item \textbf{(Crystal Structure) Prototype}: a group of crystal structures comprising the same space group or belong to enantiomorphic space group pairs and possess the same occupied Wyckoff sites (i.e.,  an \textit{isopointal} group). 
    \item \textbf{Decoratype}: an umbrella term for any subclassification of a prototype which describes the particular \textit{decoration} of the crystal structure (the following definitions are all decoratypes).
    \item \textbf{Decoratype family}: the set of all possible decoratypes within a given crystal structure prototype. The terms below inherit this terminology as well, e.g., a \textit{polaritype family} is the set of all possible polaritypes within a given prototype.
    \item \textbf{Polaritype} (binary): group of crystals within a prototype which possess the same mapping between \textit{ionic polarities} and inequivalent Wyckoff sites.
    \item \textbf{Oxitype} (scalar): group of crystals within a prototype which possess the same mapping between \textit{oxidation states} and inequivalent Wyckoff sites.
    \item \textbf{Isingtype} (binary): group of crystals within a prototype which possess the same mapping between \textit{Ising spin} and inequivalent Wyckoff sites.
    \item \textbf{Heisentype} (scalar): group of crystals within a prototype which possess the same mapping between \textit{Heisenberg spin} and inequivalent Wyckoff sites.
    \item \textbf{Momentype} (vector): group of crystals within a prototype which possess the same mapping between \textit{magnetic moment} and inequivalent Wyckoff sites. Here the magnitude of the vector would represent the spin and the direction would represent the spin vector, e.g., [0, 0, 5/2] for 5/2 $\mu_B$ along the $c$ axis. Higher-order spin or charge multipoles would require higher-dimensional representations \cite{PhysRevB.104.054412}. While not explicitly described in this work, such representations are compatible with the decoratype framework and may be of interest in future work \cite{PhysRevResearch.5.L042018,PhysRevX.14.011019}.
    \item \textbf{Ionotype} (categorical): group of crystals within a prototype which possess the same mapping between \textit{ions} and inequivalent Wyckoff sites.
    \item \textbf{Elemtype} (categorical): group of crystals within a prototype which possess the same mapping between \textit{element} and inequivalent Wyckoff sites. Differs from ionotype only in special cases, e.g., charge ordering.
    \item \textbf{Written Representation}: Building on the AFLOW Prototype notation, we add another term after the occupied Wyckoff sites consisting of the values of the associated property ordered to reflect their mapping to the corresponding Wyckoff sites. The terms in this representation (joined by underscores) are the following:
    \begin{itemize}
        \item $\langle$Anonymized formula$\rangle$
        \item $\langle$Pearson symbol$\rangle$
        \item $\langle$Space group number$\rangle$
        \item $\langle$Occupied Wyckoff site 1$\rangle$\blank$\langle$Occupied Wyckoff site 2 $\rangle$\blank...
        \item $\langle$Property value at site 1$\rangle$\blank$\langle$Property value at site 2$\rangle$\blank...
    \end{itemize}
    For example, the oxitype for a 2:4 cubic perovskite is expressed as 
    \text{AB3C\blank{}cP5\blank{}221\blank{}a\blank{}c\blank{}b\blank{}\blank{}2+\blank{}2-\blank{}4+}
    . 
\end{itemize}

\section{Data Set}

For this work, we used an expanded version of the data set from a previous work \cite{miller2023}, which draws from three crystallographic databases: the Materials Project (MP) \cite{jain2013, ong2015}, the Inorganic Crystal Structure Database (ICSD) \cite{bergerhoff1987}, and the Open Quantum Materials Database (OQMD) \cite{saal2013, kirklin2015}.
The MP and OQMD data sets include binary through quintenary compounds reported as synthesized for which all components can be assigned a meaningful integer oxidation state.
The ICSD data set includes binary and ternary compounds containing at least one transition, alkali, or alkaline earth metal  as well as at least one of the following anions: N, O, F, P, S, Cl, As, Se, Br, Te, I. 
These structures are all reported in the ICSD with the following conditions: $283 \text{ K} \leq T \leq 303 \text{ K}$ and $0.9\text{ atm} \leq P \leq 1.1\text{ atm}$.

The data sets were merged, and any non-unique compounds, that is with same isopointal group and same species, were filtered out, preferring the ICSD entry with the highest collection code or, if no matching ICSD entry was available, the MP entry with the lowest energy above the convex hull.
This process yielded 38,049 total unique compounds.
\autoref{eda} shows an overview of the chemical makeup of the MP data set, revealing a strong bias toward ternary and quaternary oxides due to database composition and synthesis trends. 

\section{Computational Methods}

\subsection{Feature Selection and Model Training}

SHapley Additive exPlanations (SHAP) is a statistical approach to quantifying the contribution of individual features towards a model's prediction \cite{lundberg2017}. SHAP values represent the contribution of each feature to the difference between the actual and baseline predictions. Here, we apply SHAP feature importance analysis for feature selection and revealing physical insights captured by our model. 

For feature selection, since the particular ordering of SHAP-derived importances is crucial to producing an accurate model, we use two tree-based models (DecisionTreeClassifier and RandomForestClassifier from scikit-learn \cite{pedregosa2011}) and TreeExplainer from the SHAP Python library which properly treats codependent features \cite{janzing2020}. 
At each step in the active learning loop, we perform feature selection prior to training the final model used to make the candidate recommendations. Using stratified 5-fold cross-validation we calculate 5 sets of SHAP values for each of the two feature selection models. Net feature importances are computed as the average SHAP values across each of the 5 folds for each of the feature selection models. 
The top 20 features based on these averaged SHAP values are then used as the feature set for each  model in the recommender committee (DecisionTreeClassifier, RandomForestClassifier, ExtraTreesClassifier, and GradientBoostingClassifier). Each of these models in the committee is trained on the entire set of available known data at that active learning step and the evenly weighted average of the committee's predictions (of the likelihood of formation) for each candidate compound are used as the candidate score. This use of a recommender committee, mostly containing ensemble models, represents a two-layered approach to avoid overfitting, a crucial aspect of creating a model designed to discover new compounds. 

In all cases, the machine learning models were instantiated with scikit-learn defaults for all parameters except class weights. Further details about default model parameters can be found in the Supplementary Materials (SM) \cite{supp}). To compensate for class imbalance, class weights were set to reflect the proportions of labels present in the training data set. Since gradient boosting tends to handle class imbalance well by construction, no class weighting adjustments were used for that model.

Section V.B and \autoref{shap} provide a detailed discussion and visualization of the feature importance in the final recommender committee.

\subsection{Polaritype Classification Workflow}

We use Pymatgen (Python Materials Genomics), an open-source materials analysis Python library, to manipulate crystal structures, analyze atomic distances, and guess oxidation states \cite{ong2013}. We use Matminer \cite{ward2018} for featurization and scikit-learn for machine learning. 
For the polaritype identification workflow, we use the structure prototype encyclopedia from AFLOW \cite{mehl2017, hicks2019} and the AFLOW-XtalFinder code \cite{hicks2021} in concert with Pymatgen to process structures and sort them into isopointal groups.
The classification workflow consists of the following steps:
\begin{itemize}

    \item[(a)] Determine the prototypes of the structures and then rewrite the files in order to render all structures of a given prototype uniform in terms of their choice of crystallographic origin, orientation, and Wyckoff labeling. \textbf{Tool}: AFLOW prototype encyclopedia \cite{mehl2017, hicks2019, hicks2021} and AFLOW-XtalFinder \cite{hicks2021}.
    
    \item[(b)] Group structures by unique combinations of sorted, anonymized chemical formula, Pearson symbol, and space group number, e.g., \texttt{ABC3\blank{}cP5\blank{}221} for the group containing cubic perovskites (to accelerate the comparison step). This pre-processing step offers significant speedup of the AFLOW-XtalFinder \texttt{compare\_structures} command, which involves pairwise comparisons.
    \textbf{Tool}: In-house code. 
    
    \item[(c)] Within each outer group created in the previous step, group \textit{isopointal} structures, or structures which belong to the same space group or enantiomorphic space group pair and can possess the same occupied Wyckoff positions when the appropriate unit cells are selected. Further details on this process can be found in the SM in \cite{supp}. \textbf{Tool}: AFLOW-XtalFinder's \texttt{compare\_structures} function. 
    
    \item[(d)] Evaluate the oxidation states of all atoms in each structure. \textbf{Tool}: Pymatgen's \texttt{ValenceIonicRadiusEvaluator} class. 
    
    \item[(e)] Create ion--Wyckoff site mappings within each group of isopointal structures.
    \textbf{Tool}: In-house code.
    
    \item[(f)] Collapse equivalent ion--Wyckoff site mappings. Each unique mapping comprises a single polaritype. 
    \textbf{Tool}: In-house code and AFLOW-XtalFinder's \texttt{get\_unique\_atom\_decorations} function.
\end{itemize}

\subsection{First-Principles Calculations}

We performed first-principles density functional theory (DFT) calculations using the Vienna \textit{Ab initio} Simulation Package (VASP) \cite{kresse1996} with projector-augmented-wave pseudopotentials \cite{blochl1994, kresse1999} within the Perdew-Burke-Ernzerhof generalized gradient approximation (PBE) \cite{perdew1996}. Calculations including the Hubbard $U$ correction used the method introduced by Dudarev et al.\ \cite{dudarev1998}, which features $U_{\mathrm{eff}}$ as the only parameter.
We used the Phonopy package with a $2\times2\times2$ supercell and 0.01 \AA\ displacements for pre- and post-processing of the dynamical matrix \cite{togo2015}. 
Phonon spectra were calculated using the approach by Gonze et al.\ for the non-analytical corrections to capture the splitting of the longitudinal optical (LO) and transverse optical (TO) modes \cite{gonze1994, gonze1997}.
We parsed and visualized our results with  Pymatgen \cite{ong2013}.

\subsubsection{High-Throughput Stability Screening}
To screen candidates for stability, we perform convex hull calculations using structures from the Materials Project and the OQMD. 
VASP input and output files can be found in the associated Dryad entry \cite{supp_dryad}.
Compounds within 26\,meV/atom of the convex hull are considered ``stable'' for the purpose of this study.
More specifically, for a candidate with formula \ce{A4B2C}, we perform the following procedure to obtain the convex hull: 
\begin{itemize}
    \item[(a)]  Gather all structures from OQMD and MP within the A-B-C composition space (including pure elements and binaries).
    \item[(b)] Limit the structures to those within 26\,meV/atom of the convex hull.
    \item[(c)] Group the structures by prototype using AFLOW-XtalFinder's. \texttt{compare\_structures} function and remove duplicate structures, prioritizing MP structures with the largest MP ID. Full command details can be found in  \cite{supp}. 
    \item[(d)] Using VASP, fully relax each structure in four successive runs with \texttt{ISIF} taking the following values, in order: 6, 2, 3.
    \item[(e)] Perform self-consistent field calculations on the relaxed structures to obtain the total energy.
    \item[(f)] Cache calculation results for unary and binary compounds to prevent duplicate calculations for future candidates.
    \item[(g)] Use Pymatgen's \texttt{PhaseDiagram} class to calculate the convex hull.
\end{itemize}
VASP settings for these types of calculations are modified versions of Pymatgen's MPRelaxSet and MPStatic set. 
To avoid incompatible comparisons across compositions on the same phase diagram, the Hubbard correction was not used for the convex hull calculations, only for the high-fidelity calculations to calculate relevant properties of promising candidate compounds.
Details are provided in \cite {supp}. 

\begin{figure}
    \centering
    \includegraphics[width=0.7\linewidth]{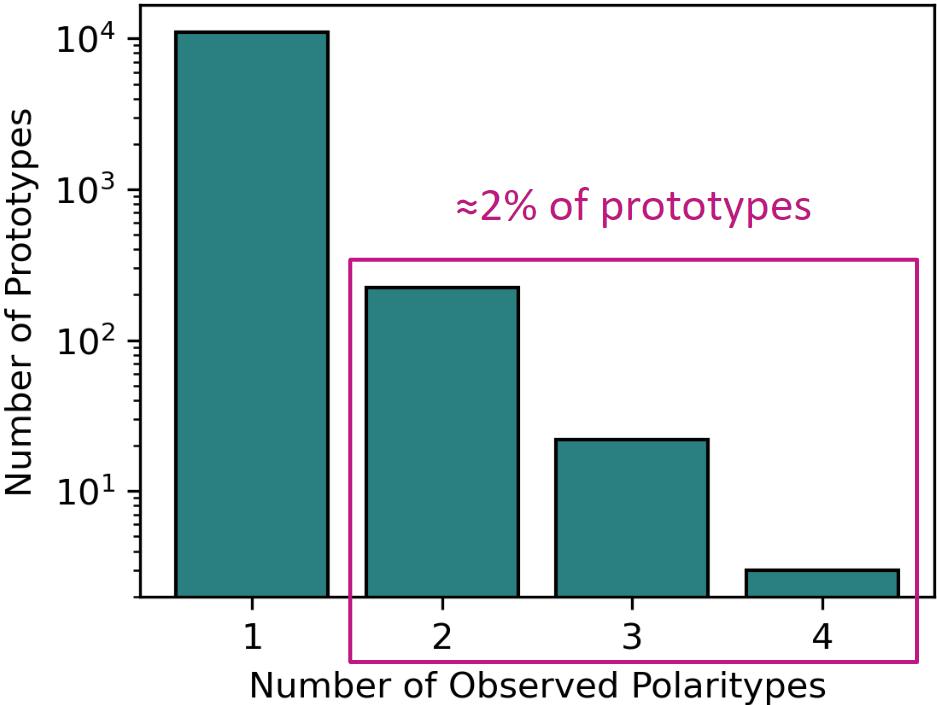}
    \caption{Histogram showing the number of observed polaritypes in each structure prototype observed in the data set. Prototypes with more than one observed polaritype make up $\approx$2\% of the population.}
    \label{pop_npolaritypes}
\end{figure}

\subsubsection{High-Fidelity Simulations}

To simulate thin films of the anti-RP candidates, a representative layer, corresponding to one rock salt and one perovskite layer, is first sliced from the fully relaxed bulk structure. $N$ layers are stacked with the appropriate in-plane shift applied to every other layer to achieve an $N$ layer thin film. To maintain centrosymmetry of the nonpolar phase, we only simulate thin films where $N$ is odd and we remove one rock salt layer from the bottom of the film, meaning anti-perovskite layers terminate each surface. Finally, we add 20\,\AA\ of vacuum to avoid interactions between the periodic images (in addition to use of VASP's dipole correction feature: \texttt{LDIPOL}).

\section{Results and Discussion}

\subsection{Population Statistics}

\begin{figure}[b]
    \centering
    \includegraphics[width=0.9\linewidth]{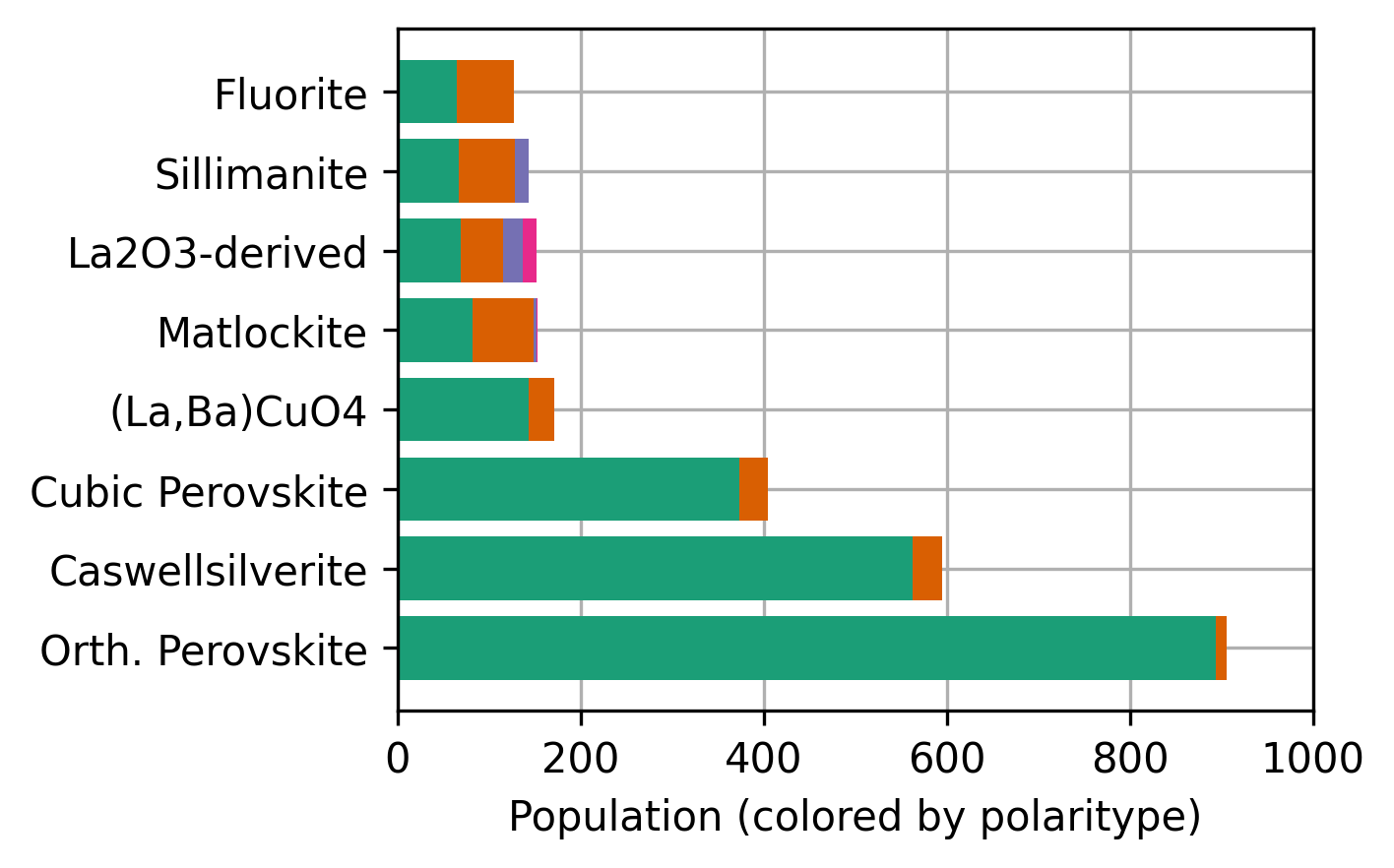}
    \caption{Polaritype populations for the top 8 most populous prototypes with at least 5 compounds observed in a nondominant polaritype. Common names on the y-axis were are those assigned by Robocrystallographer \cite{ganose2019}. (La,Ba)CuO$_4$ is the same anti-Ruddlesden-Popper structure prototype containing the polaritype of interest in this work.}
    \label{pop_top_protos}
\end{figure}

In the data set with 38,049 total compounds, representing 11,306 structural prototypes, we observed known compounds representing 11,583 unique polaritypes. However, we calculated the total number of possible polaritypes as 132,115, meaning that only 8.8\% of theoretically possible charge polarity decorations of the lattices in the data set have been observed in experiment. 
Of the 11,306 prototypes, only 249 (2.2\%) possess polaritype families with more than one observed polaritype (\autoref{pop_npolaritypes}). 
We find that the vast majority of crystal structures (97.8\%) can only host a single stable cation/anion arrangement, consistent with the fact that anti-structures (or nontrivial polaritype families) are relatively rare.
Some structure prototypes like rock salt possess only one distinct polaritype by symmetry, as interchanging the cation and anion positions corresponds to a trivial translation, 
but these comprise only 3.8\% of the total observed structure prototypes.

\begin{figure}[b]
    \centering
    \includegraphics[width=\linewidth]{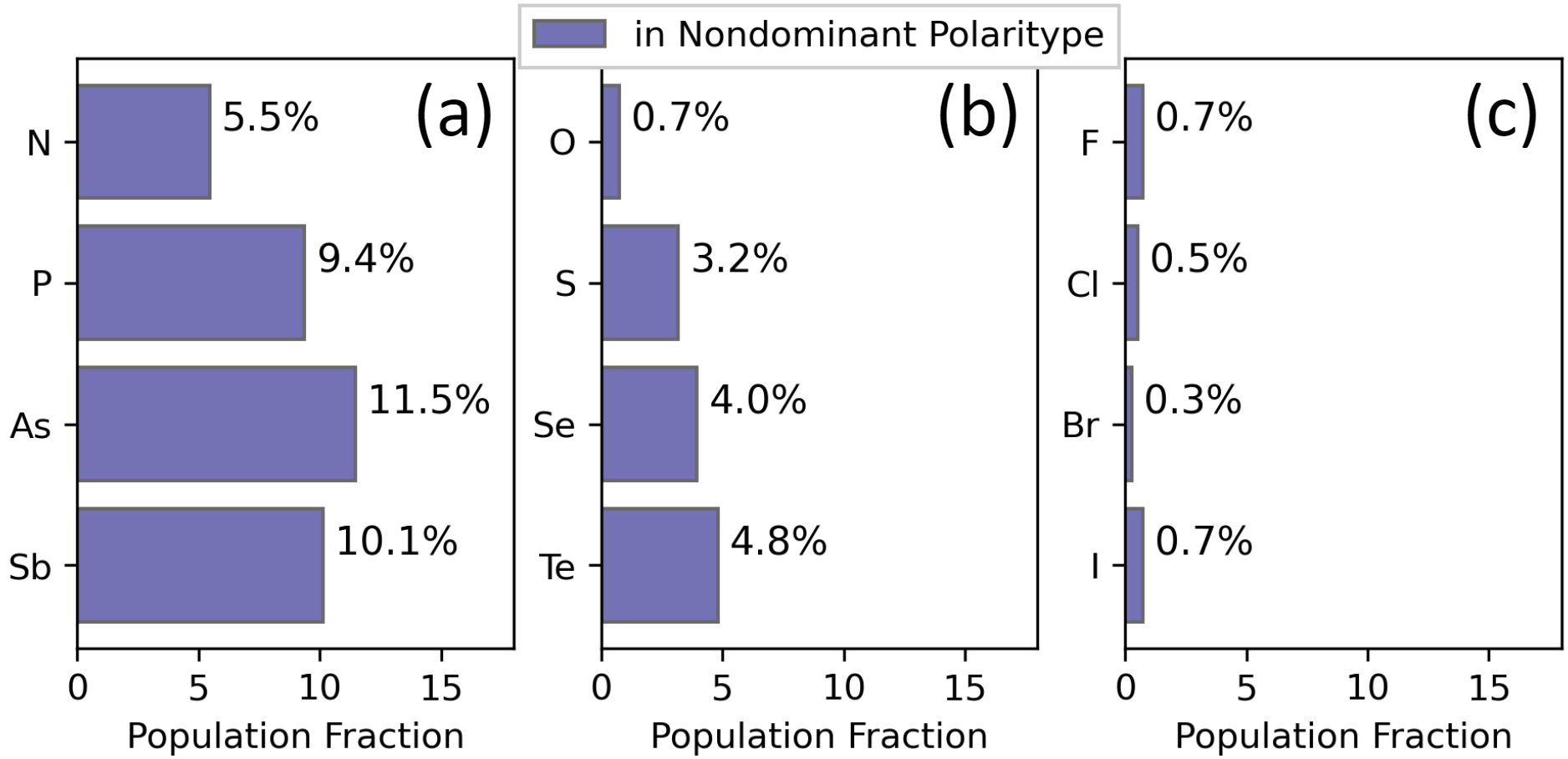}
    \caption{Anion-separated bar chart showing the percentages of compounds belonging to nondominant polaritypes, i.e., possessing a Wyckoff site--ion polarity mapping that is \textit{not} the most common within the respective structural prototype. Each bar represents the nondominant polaritype percentage found in the population of all compounds containing an anion of the specified element.}
    \label{pop_nondom}
\end{figure}

\begin{figure*}
    \centering
    \includegraphics[width=0.64\linewidth]{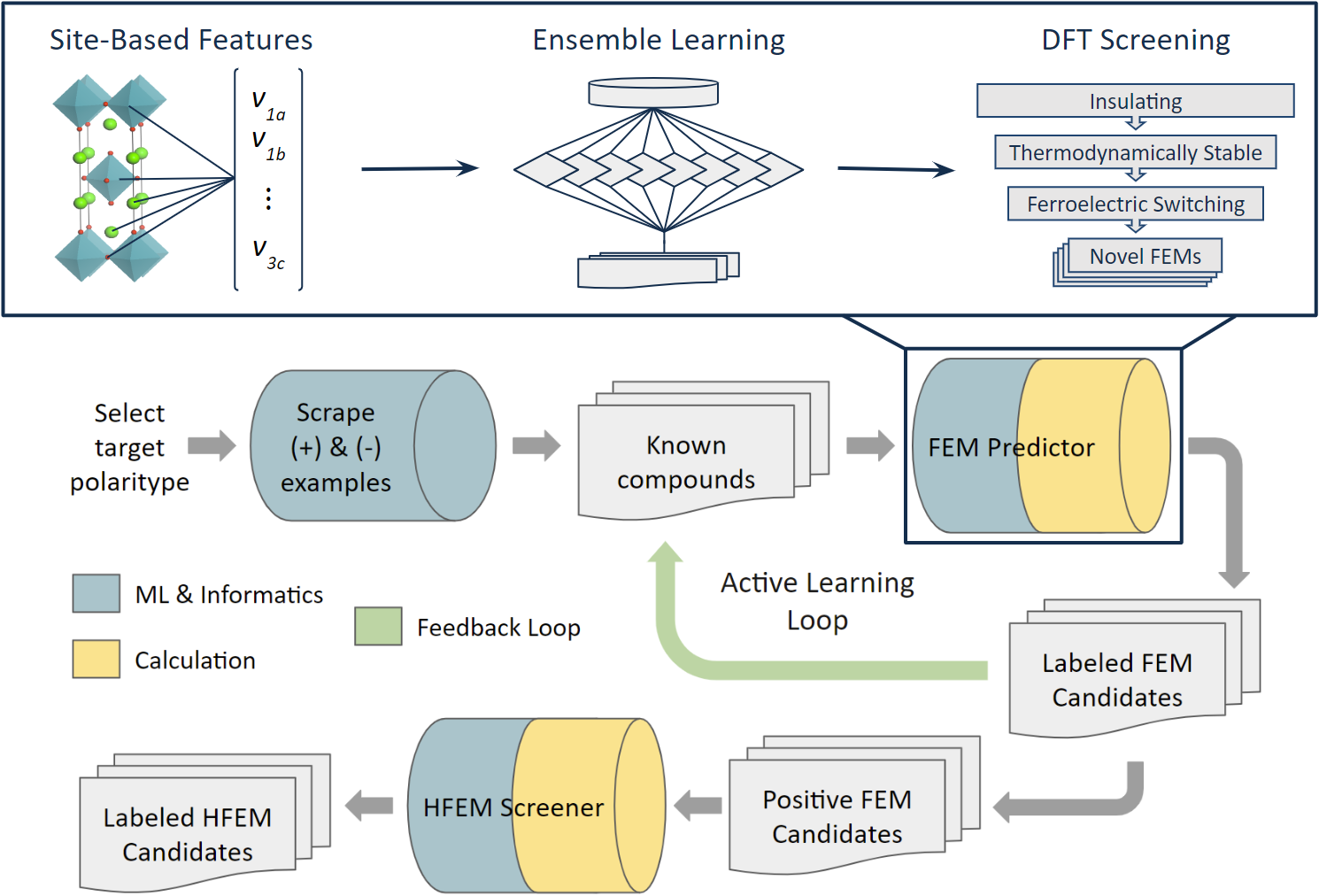}
    \caption{Schematic of the (H)FEM discovery workflow. The \textit{FEM Predictor} consists of three separate steps: featurization, machine learning prediction of formation within the selected polaritype, and DFT screening for ferroelectric behavior. The \textit{HFEM Screener} consists of phonon calculations with the non-analytical correction to calculate the frequencies of the longitudinal optical (LO) and transverse optical (TO) modes. If and only if the LO and TO modes are both unstable, i.e., exhibit imaginary frequencies, and there are no other unstable modes, then the crystal is considered hyperferroelectric. 
    }
    \label{fig:workflow}
\end{figure*}

Most of the populous nontrivial prototypes contain only two distinct observed polaritypes and are dominated by a single polaritype (\autoref{pop_top_protos}), cases in which the ``structure/anti-structure'' terminology suffices. The vast majority of known prototypes have only a single known polaritype, indicating that site flexibility is not a common characteristic among structural prototypes. 
In the rare instances where a structure accommodates multiple polaritypes, e.g., La$_2$O$_3$-derived compounds as shown in \autoref{pop_top_protos}, we observe few unifying features, aside from the rarity of face-sharing connectivity and the prevalence of layered crystal habits.
These observations reinforce the hypothesis that increased structural rigidity generally suppresses the diversity of accessible polaritypes.

Furthermore, we see familiar periodic trends represented in which compounds are found in nontrivial polaritype families. \autoref{pop_nondom} shows that the occurrence of anions in nondominant polaritype families is heavily weighted toward less electronegative members. 
Moving from panels (a) to (c) in \autoref{pop_nondom} corresponds to moving rightward in the periodic table, which results in a significant decrease in the population of nondominant polaritype compounds, i.e., those currently referred to as anti-structures. 
These percentages are obtained by summing the minority polaritypes and dividing by the total population of compounds containing a certain anion.
These compounds, which contain anions from that particular column, show a drastic reduction, with pnictides being approximately an order of magnitude more prevalent than halides.

We attribute the trends appearing in \autoref{pop_nondom} to the different electronegativities of the anions. Strongly electronegative anions, being small in size, form short and rigid ionic bonds, which leave little room for structural flexibility. As a result, they adopt structures that are less capable of accommodating ions of significantly different sizes. In contrast, less electronegative anions, which tend to bond with cations of similar size, create more flexible structures. When the size disparity between the cations and anions is large, the resulting structural prototypes experience significant strain, making ion substitution difficult. However, structural prototypes formed by ions with similar sizes are less prone to strain and can more easily accommodate substitutions without disrupting the stability of the structure.

\begin{figure}
    \centering
    \includegraphics[width=0.39\linewidth]{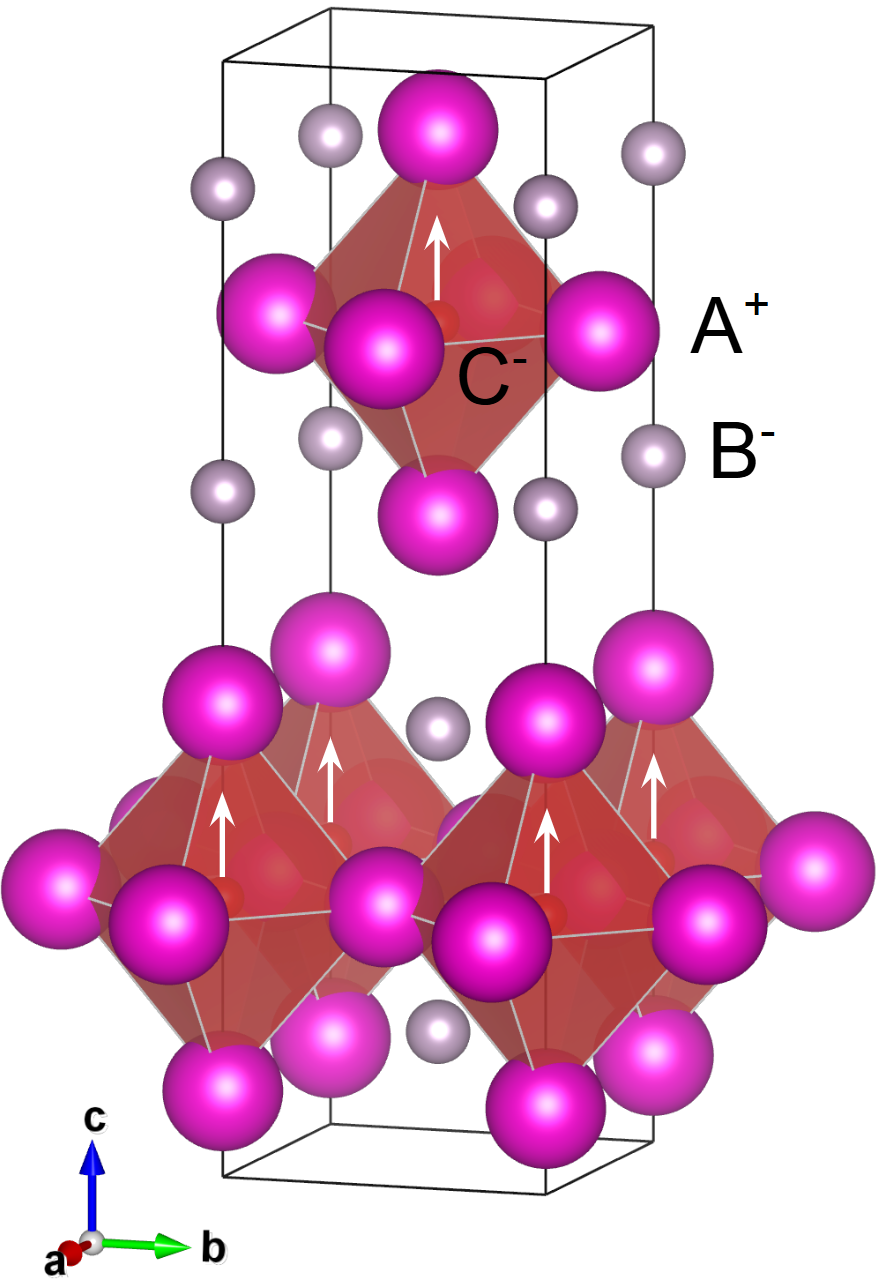}
    \caption{Schematic of the A4B2C\blank{}tI14\blank{}139\blank{}e\blank{}a\blank{}ce\blank{}\blank{}+\blank{}-\blank{}- polaritype where the A cation forms an octahedron containing a C anion (CA$_6$) and the B anion sits in the voids between octahedra.
    White arrows indicate polar displacement directions of the oxide (C$^-$) anions, which we observe in some of the oxide members of the 
    family (where $\text{C}=\text{O}$).
    }
    \label{fig:struc-antirp}
\end{figure}

\subsection{Use Case: 
Anti-Ruddlesden-Popper (Hyper)Ferroelectric Materials Discovery} 

Existing methods for discovery of ferroelectric materials (FEMs) and hyperferroelectric materials (HFEMs) are either too expensive to apply broadly \cite{markov2021}, too broad to capture much of our existing physical and chemical understanding \cite{ricci2024, smidt2020}, or are specific to a particular subsets of materials space \cite{zhang2020, kruse2023}. Since FEMs are rare, we must simultaneously reduce the screening cost and leverage our domain knowledge to accelerate discovery. To that end, we built a faster, more targeted FEM screening workflow (\autoref{fig:workflow}). It builds on the polaritype classification workflow introduced earlier, thereby taking  advantage of existing materials property databases. The workflow further employs an active learning framework to continuously update the recommender model with new information obtained via both in-situ and in-silico methods.

Novel developments in the workflow include: (i) phase stability prediction 
within a ferroelectricity-prone family, (ii) iterative updates to the model with the DFT results from the previous generation of candidates, (iii) evaluation of strain effects on the ferroelectric mode, and (iv) HFEM screening of FEM candidates. 

As a test case for this workflow, we selected the perovskite-derived anti-Ruddlesden-Popper prototype, members of which have been predicted to be ferroelectric (\ce{Eu4Sb2O}, \ce{Ba4As2O}) and hyperferroelectric (\ce{Ba4Sb2O}) \cite{markov2021, kang2020}.
In addition to inheriting ferroelectric potential from their perovskite parent structure, layered perovskite-derived structures also allow coupling between ferroic distortions and other lattice distortions or physical properties \cite{benedek2015} and provide additional chemical tuning with which to stabilize noncentrosymmetric phases \cite{balachandran2014}.
For example, Markov and colleagues predicted coupled ferromagnetism and ferroelectricity in the anti-Ruddlesden-Popper\ce{Ba4Sb2O} \cite{markov2021} while Balachandran and colleagues showed that cation ordering can be used to lift inversion symmetry in $n=1$ Ruddlesden-Popper oxides. 

In particular, we selected the A4B2C\blank{}tI14\blank{}139\blank{}e\blank{}a\blank{}ce\blank{}\blank{}+\blank{}-\blank{}- polaritype as our search space, made promising by the predicted ferroelectric members \ce{Eu4Sb2O}, \ce{Ba4As2O}, and \ce{Ba4Sb2O}.
With this workflow, we predicted six novel FEM candidates, three of which have strain-activated HFE potential, in the $n=1$ anti-Ruddlesden-Popper family (with the structure in \autoref{fig:struc-antirp}).

\subsubsection{Polaritype Prediction Workflow: Model preparation}
\begin{itemize}
    \item[(a)] Select a single target polaritype.
    \item[(b)] Gather training data.
    \begin{itemize}[leftmargin=0.5cm]
        \item[-] Positive examples: Structures stable in the target polaritype. Stable is defined as being observed experimentally with $283 \text{ K} \leq T \leq 303 \text{ K}$ and $0.9\text{ atm} \leq P \leq 1.1\text{ atm}$ or having a calculated energy above hull less than 25\,meV/atom.
        \item[-] Negative examples: (i) Structures stable in (i) the associated structure prototype but as a different polaritype and  structures stable in (ii) a different structure prototype
    \end{itemize}
    \item[(c)] Compute the Wyckoff site-specific element-based features (described in-depth in the SM \cite{supp}) and use mean value imputation to fill in missing data.
    \item[(d)] Train a polaritype classifier (committee of four tree-based classifiers) to predict whether compositions are stable in the selected polaritype. This classifier will serve as the ``recommender model'' in the active learning discovery workflow.
\end{itemize}

\begin{figure}
    \centering
    \includegraphics[width=0.95\linewidth]{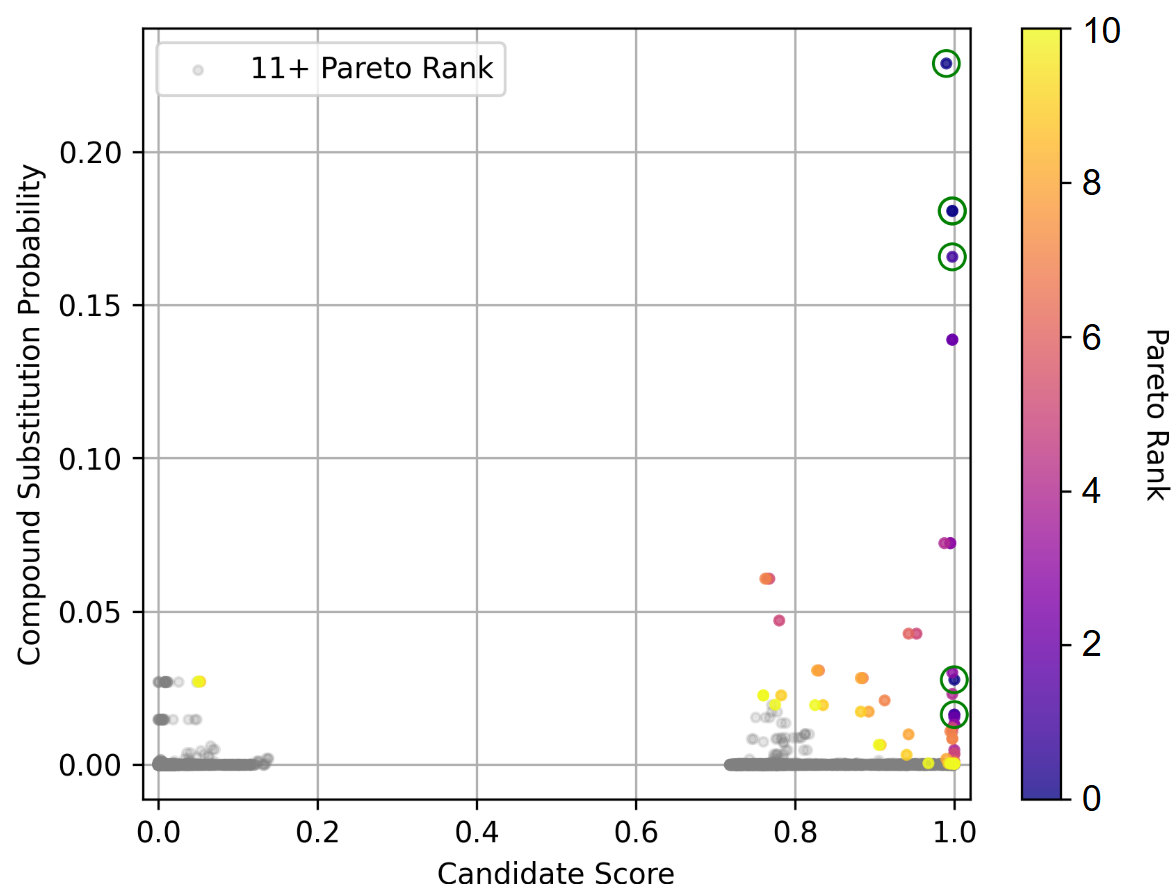}
    \caption{Visualization of the candidate selection process that occurs at each active learning step. Points represent candidate compounds plotted by their score assigned by the polaritype classifier (likelihood of exhibiting the A4B2C\blank{}tI14\blank{}139\blank{}e\blank{}a\blank{}ce\blank{}\blank{}+\blank{}-\blank{}- polaritype) and the compound substitution probability (CSP) relative to the most similar known compound. Points are colored by their Pareto rank with low ranks representing those which dominate all points of higher rank along each axis. Green circles indicate the lowest Pareto-ranked compounds, those selected for simulation. In the first active learning step of the 6-candidate-per-batch model, \ce{Rb4Cl2O}, \ce{K4Cl2O}*, \ce{Na4Cl2O}*, \ce{Yb4Bi2O}*, \ce{Sm4Sb2O}, and \ce{Cs4Br2O}, were recommended and those with * were predicted stable.}
    \label{fig:Pareto}
\end{figure}

 \begin{figure*}
    \centering
    \includegraphics[width=0.85\linewidth]{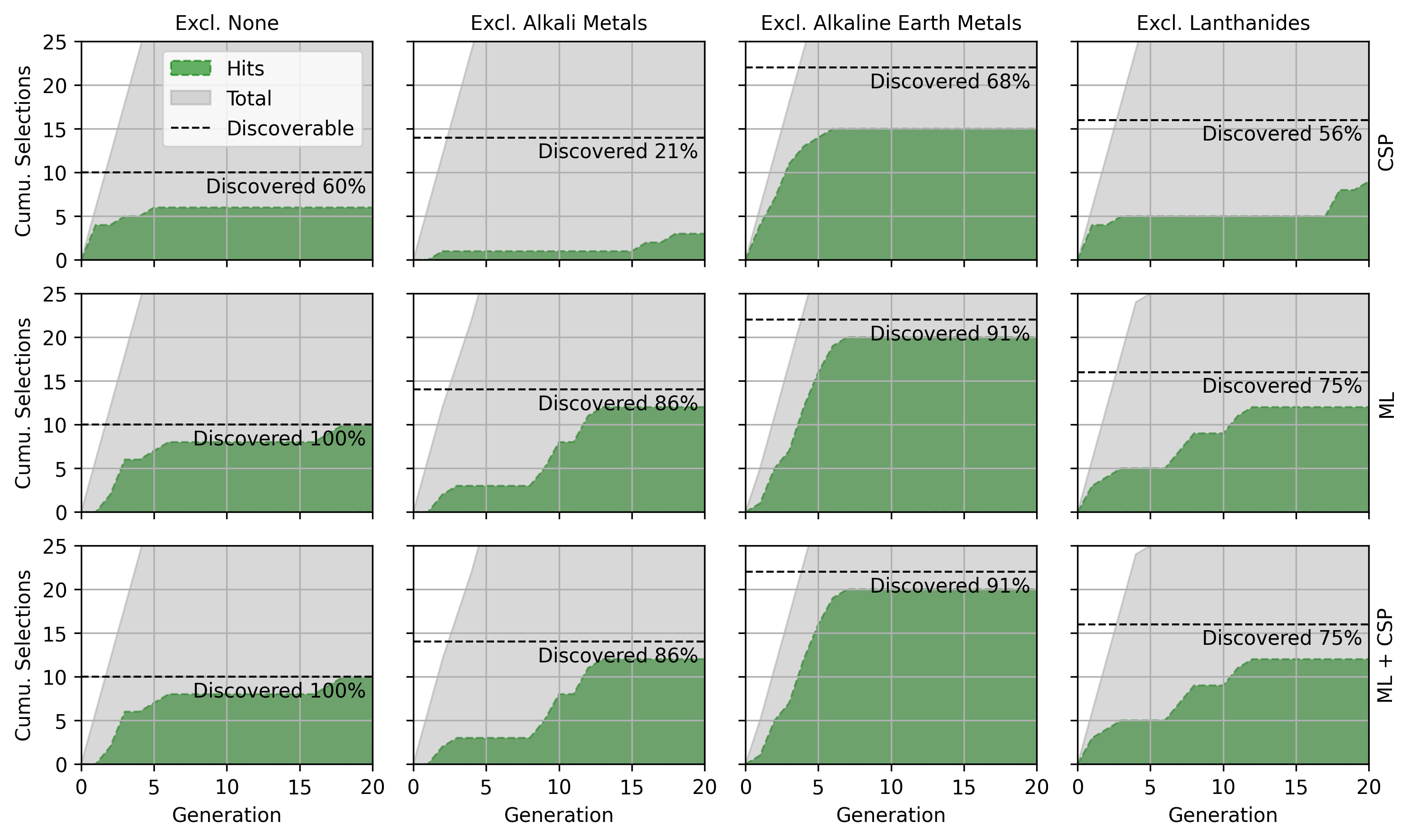}
    \caption{Discovery results for the A4B2C\blank{}tI14\blank{}139\blank{}e\blank{}a\blank{}ce\blank{}\blank{}+\blank{}-\blank{}- polaritype discovery workflow (with a batch size of 6) where the green area represents the number of correctly identified A4B2C\blank{}tI14\blank{}139\blank{}e\blank{}a\blank{}ce\blank{}\blank{}+\blank{}-\blank{}- compounds. Rows represent trials using different selection criteria (shown on the right of each row). Columns represent different validation data schema, excluding the specified group of examples from the training data. For example, in the first column, all available training data are used while in the second column, all training example containing alkali metals are excluded from the initial training, leaving four additional examples to be discovered -- explaining why the number of ``discoverable'' examples (represented by the horizontal dashed line) increases by four. Results for other batch sizes are shown in the SM \cite{supp}.}
    \label{fig:discovery}
\end{figure*}

\subsubsection{Polaritype Prediction Workflow: Active Learning Loop}
\begin{itemize}
    \item[(a)] Generate a list of candidate polaritype members consisting of all possible charge-balanced compositions.
    \item[(b)] For each candidate:
    \begin{itemize}[leftmargin=0.5cm]
        \item[-] Use the polaritype classifier to predict the likelihood of the candidate composition being stable in the target polaritype, which we will assign as \textit{candidate scores}.
        \item[-] Calculate the compound substitution probability (CSP), an estimate of how likely the compound is to be synthesizable.
    \end{itemize}
    \item[(c)] Generate a plot of CSPs and candidate scores (\autoref{fig:Pareto}).
    \item[(d)] Select the $N$ best candidates from the Pareto front of this plot to perform DFT-based stability and ferroelectric screening (green circles indicate selected candidates in \autoref{fig:Pareto}). This corresponds to the ``ML + CSP'' selection strategy shown in \autoref{fig:discovery}. The other two strategies use only the CSP and candidate (ML) scores for candidate selection.
    \item[(e)] Add the results to the training data set and retrain the polaritype classifier.
    \item[(f)] Repeat steps (a)-(e) until all available resources are spent, sufficient candidates have been discovered, or all remaining candidates fail to meet the selected threshold of likelihood to be stable in the target polaritype.
    
\end{itemize}

\subsubsection{Results and Discussion}
Numerous studies have highlighted the challenge of developing models that generalize effectively across diverse chemical families. To assess this generalizability, we conduct a series of comparative active learning trials in which specific chemical families are systematically excluded from the training data. This approach allows us to evaluate the model’s ability to ``rediscover'' the withheld compounds, thereby testing its capacity to extrapolate beyond its training distribution.

\begin{figure}
    \centering
    \includegraphics[width=0.95\linewidth]{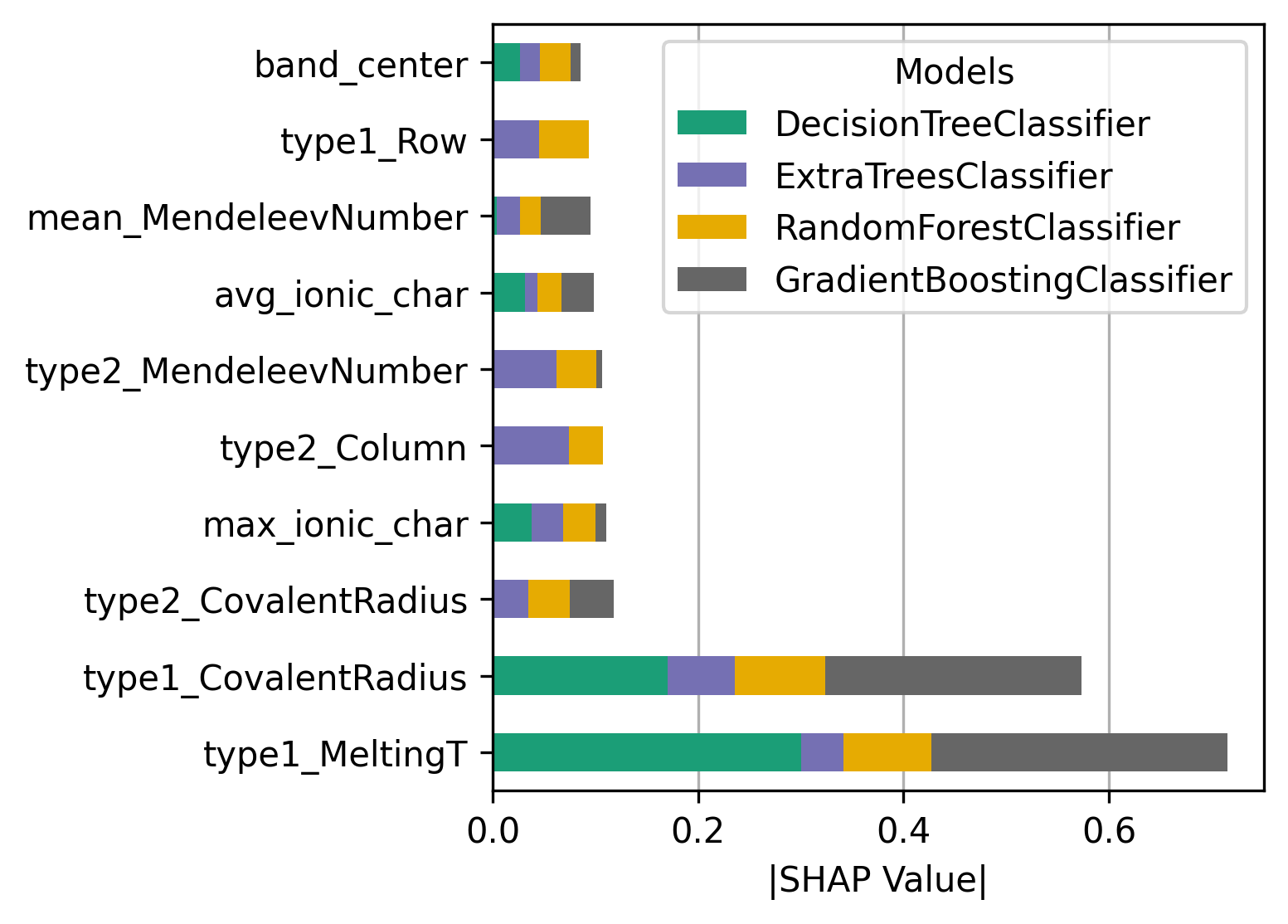}\vspace{-8pt}
    \caption{Top 10 absolute SHAP values averaged across 10 generations of active learning for each of the 4 classifiers used (as a committee) for feature reduction before training the polaritype classifier (random forest). The \texttt{typeX\_} prefix to feature names indicates the value describes the ion at a specific Wyckoff site in the A4B2C\blank{}tI14\blank{}139\blank{}e\blank{}a\blank{}ce\blank{}\blank{}+\blank{}-\blank{}- structure, specifically types 0, 1, and 2 correspond to B, C, and A, respectively.}
    \label{shap}
\end{figure}

\begin{figure*}
    \centering
    \includegraphics[width=0.97\linewidth]{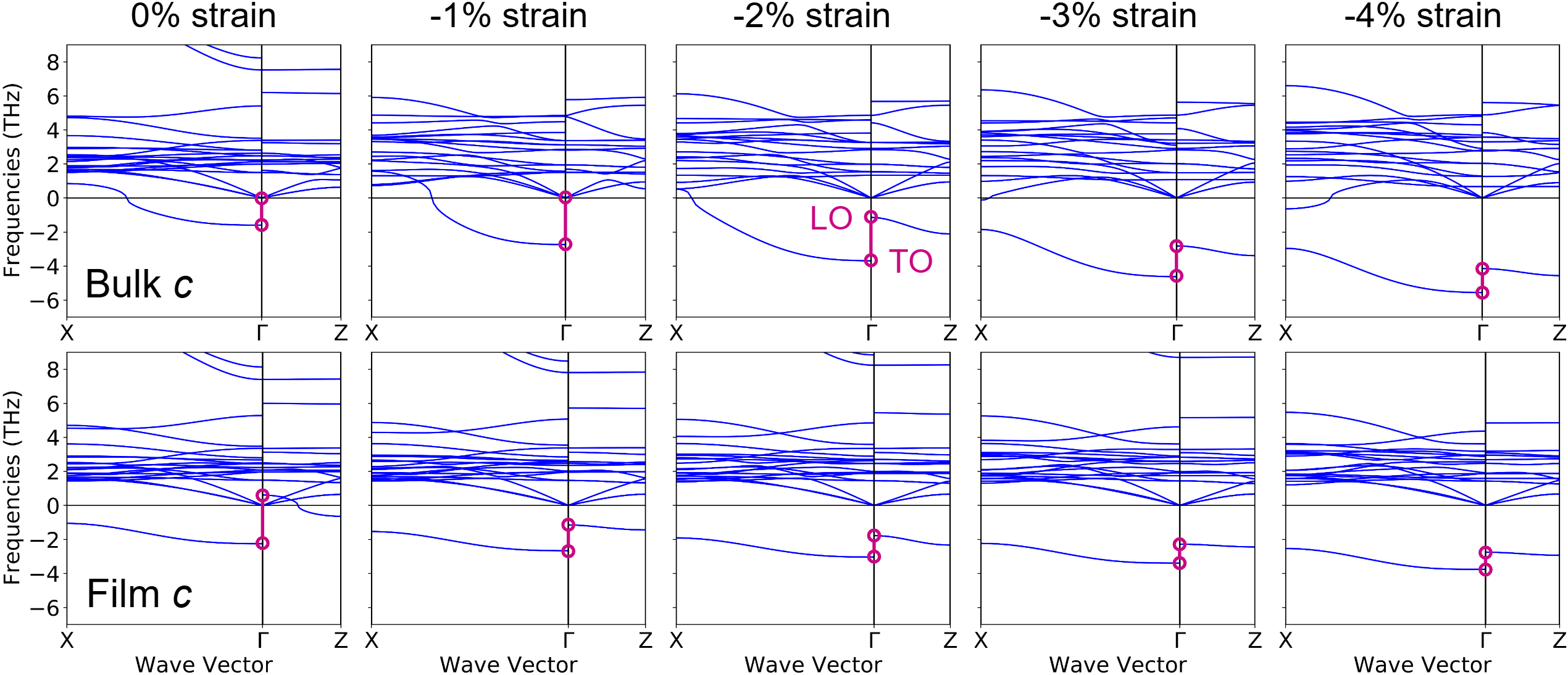}
    \caption{Phonon dispersions (including the non-analytical corrections \cite{gonze1994,gonze1997}) at various levels of epitaxial strain for \ce{Eu4Bi2O} with the $c$ lattice parameter as determined by relaxation in a (top) bulk or (bottom) film structure. Longitudinal optical (LO) and transverse optical (TO) modes are indicated in pink. The onset of hyperferroelectricity occurs near 2\% strain for the bulk $c$ structure and -1\% strain for the film $c$ structure. Negative mode values indicate imaginary frequencies.}
    \label{fig:phon-strain}
\end{figure*}

\autoref{fig:discovery} summarizes the results of these studies. 
Each plot shows the cumulative number of compounds probed (grey) and number of true positives discovered (green) relative to the total discoverable (those eventually predicted stable by our DFT calculations) candidates in the data set (dashed line). 
Each row of plots represents a different active learning selection strategy: (from top to bottom) CSP, the machine learning (ML) model developed in this work, and a evenly weighted combination of both, as shown by the Pareto front in \autoref{fig:Pareto}.
The columns in the figure represent various data partitioning strategies. The headings of each column name the group of compounds excluded from the training data. For example, ``Excl.\ Alkali Metals'' indicates that the model was trained solely on known examples (positive and negative) that do not include alkali metals, while all alkali metal-containing compounds were placed in the testing set. In this setup, the positive examples are considered discoverable compounds, explaining the variation in the dashed line across columns. This partitioning approach allows us to evaluate the generalizability of the discovery workflow by evaluating it in scenarios where entire families of compounds are unknown to the model.
With this approach, our models demonstrate generalization capability, with 75-91\% of out-of-distribution stable compounds discovered when a chemical family is excluded from the training data.

In the first column, where no previously reported compounds are excluded from the training data, we find that while the CSP strategy misses 4 of the novel compounds,  the ML model discovers all 10 novel stable compounds within 18 generations (or after examining 108 out of 606 compounds). 
This constitutes a 9\% discovery rate with 100\% coverage, where random guessing would average a 1.6\% discovery rate, and the CSP strategy achieved only 5\% discovery rate with 60\% coverage.
Comparing the three rows, we find that our ML model consistently outperforms the CSP metric. Additionally, the combination of CSP strategy and the ML model performs the same as the ML model alone. This indicates that our model is learning both broad chemical co-occurrence trends, as encoded in the CSP, and underlying chemical rules for crystal formation.

In the three rightmost columns of \autoref{fig:discovery}, which exclude compounds containing alkali metals, alkaline earth metals, and lanthanides, the observed discovery rates are 86\%, 91\%, and 75\%, respectively.
These relatively high rates, despite excluding entire segments of the periodic table, show that our ML model can extract significant chemical insight from the decoratype-based features. This allows it to rapidly explore unknown regions of the chemical space to predict novel materials.
This case study in the A4B2C\blank{}tI14\blank{}139\blank{}e\blank{}a\blank{}ce\blank{}\blank{}+\blank{}-\blank{}-  polaritype provides strong evidence that the predictive power of our decoratype-based model extends beyond interpolation within chemical families, demonstrating its ability to generalize across the periodic table.

In \autoref{shap}, we show the SHAP values for the top 10 most important features in predicting stable members of the A4B2C\blank{}tI14\blank{}139\blank{}e\blank{}a\blank{}ce\blank{}\blank{}+\blank{}-\blank{}- polaritype.
In line with insights from crystal chemistry, the most informative features are those associated with intrinsic structural and electronic characteristics that govern thermodynamic stability.
Melting temperatures (\texttt{type1\_MeltingT}) and covalent radii (\texttt{type1/type2\_CovalentRadius}) emerge as the most significant factors, with melting temperatures reflecting bond strength, while covalent radii influence the atomic packing and bond formation. 
The two ionic character features (\texttt{max/avg\_ionic\_char}) relate to the charge transfer and bonding tendencies of the ions, both of which play a crucial role in crystal stabilization. 
Mendeleev number-based features (\texttt{mean/type\_MendeleevNumber}) help capture periodic trends that correlate with chemical reactivity and stability. The row and column features (\texttt{type1\_Row, type2\_Column}) contribute information about elemental periodicity, affecting bonding environments and electronic structure. Lastly, \texttt{band\_center} offers information on density of states contributions, influencing stability by determining the likelihood of electronic transitions and bonding strength. 
The prominence of these features suggests that the polaritype classifier successfully encodes the influence of atomic-level structural characteristics and electronic attributes on the thermodynamic stability of candidate compounds.

\begin{figure}
    \centering
    \includegraphics[width=\linewidth]{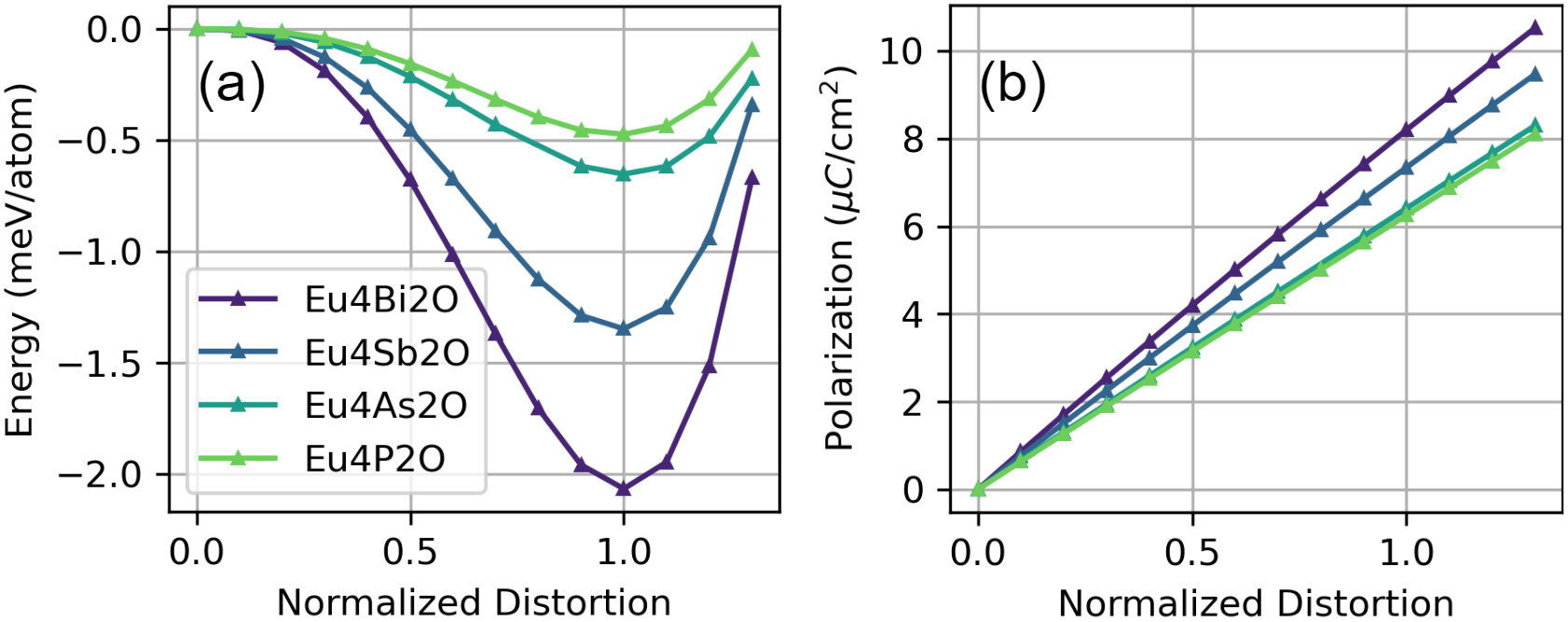}
    \caption{(a) Energy and (b) polarization of the \ce{Eu4Pn2O} (Pn = P, As, Sb, and Bi) series as interpolation between the nonpolar and polar phases. The $x$ axis represents a normalized measure of the ferroelectric displacements, with 0 representing the nonpolar reference structure and 1 representing the polar ground state structure.
    For applied strains that activate ferroelectricity, the Yb-containing analogues display qualitatively consistent trends.
    }
    \label{fig:stats-Eu-series}
\end{figure}
\subsection{Structure–Property Relationships in Discovered (Hyper)Ferroelectrics}
The new ferroelectric materials predicted by this workflow all belong to the \ce{Eu4Pn2O} and \ce{Yb4Pn2O} families 
(Pn = P, As, Sb, and Bi) represent the most interesting cases 
due to the strain-tunable ferroelectric instability. 
For the \ce{Yb4Pn2O} family, phonon calculations shown in \supf1.
found dynamically stable structures, indicating $I$4/$mmm$ paraelectric ground states.
For the \ce{Eu4Pn2O} family, all members are predicted to exhibit an unstable polar mode in the nonpolar $I$4/$mmm$ (space group 139) reference structure (\autoref{fig:phon-strain}). 
For \ce{Eu4P2O}, our calculations find that the $X$-point mode leads to a nonpolar $Cmce$ ground state structure, lower in energy than the polar structure.
The three other members of the Eu family feature a single unstable mode (\supf1), corresponding to a $I$4/$mmm$ $\rightarrow$ $I$4$mm$ ferroelectric phase transition in which the oxygen atom displaces along the $c$ axis (\autoref{fig:struc-antirp}).
Thus, we identify \ce{Eu4As2O}, \ce{Eu4Sb2O}, and \ce{Eu4Bi2O} as proper ferroelectrics.

\begin{table}
\caption{
Nonpolar reference structure, ferroelectric behavior, computed electronic band gap, and energy above the convex hull for \ce{A4B2O} (H)FE compounds. 
Key: ``$\epsilon$-ind'' indicates that ferroelectricity can be induced via epitaxial strain; 
Superscript ``${R}$'' indicates already reported in literature; \protect\cite{hoenle1998,wang1977,markov2021,baranets2022,burkhardt1998, schaal1998};
Superscript ``${H}$'' indicates that hyperferroelectricity can be induced via epitaxial strain; 
and 
``Synthesized'' indicates that hull energy was not calculated since the material was already reported synthesized.
}
\begin{ruledtabular}
\begin{tabular}{lllll}
{Formula} & {Structure} & {FE?}    & \multicolumn{1}{c}{$E_{hull}$ (meV/atom)} & \multicolumn{1}{c}{{$E_g$ (eV)}} \\ \hline
\ce{Yb4P2O}      & $I$4/$mmm$         & $\epsilon$-ind. & 0.45                                                     & 0.98                                    \\
\ce{Yb4As2O}     & $I$4/$mmm$$^{R}$         & $\epsilon$-ind. & Synthesized                                                     & 0.94                                    \\
\ce{Yb4Sb2O}     & $I$4/$mmm$$^{R}$         & $\epsilon$-ind. & Synthesized                                                     & 0.63                                    \\
\ce{Yb4Bi2O}     & $I$4/$mmm$         & $\epsilon$-ind. & 5.89                                                     & 0.50                                    \\\hline
\ce{Eu4P2O}      & $Cmce$             & PE$^{R}$             & 4.47                                                     & 0.76                                    \\
\ce{Eu4As2O}     & $I$4/$mmm$$^{R}$         & FE$^{H}$              & Synthesized                                                     & 0.77                                    \\
\ce{Eu4Sb2O}     & $I$4/$mmm$$^{R}$         & FE$^{R,H}$             & Synthesized                                                     & 0.58                                    \\
\ce{Eu4Bi2O}     & $I$4/$mmm$$^{R}$        & FE$^{H}$              & Synthesized                                                     & 0.39       \\    
\end{tabular}
\end{ruledtabular}
\label{tab:stats}
\end{table}

\begin{figure}
    \centering
    \includegraphics[width=\linewidth]{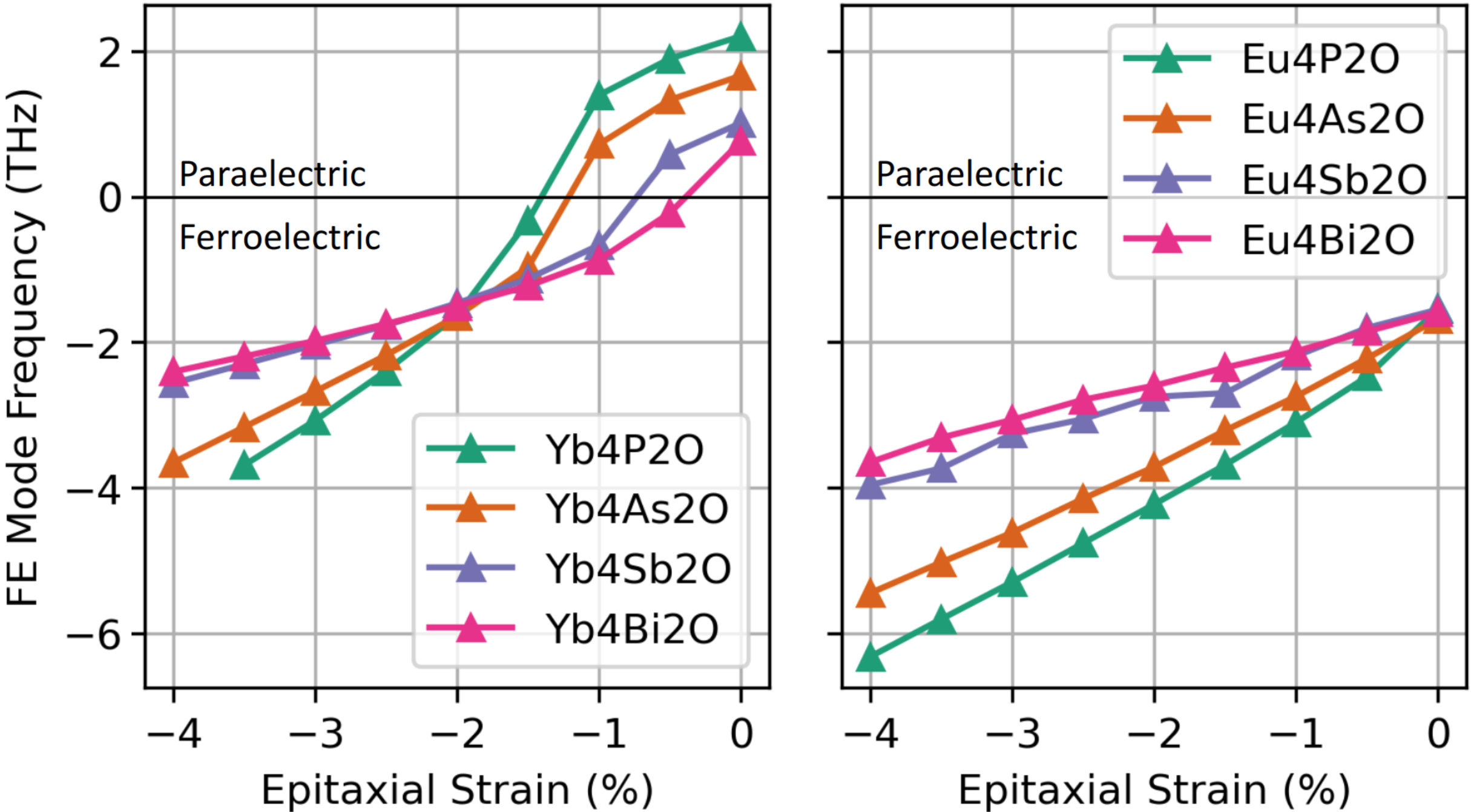}
    \caption{Frequency of the ferroelectric mode, specifically the transverse optical mode, across strain and chemistry. All members of the Yb series are predicted to be paraelectric with ferroelectric phases accessible via compressive strain. All members of the Eu series are predicted ferroelectric, except \ce{Eu4P2O} which exhibits a different, nonpolar ground state structure despite the presence of the polar instability in the high-symmetry reference structure. Negative mode values indicate imaginary frequencies.
    }
    \label{fig:fe-modes-strain}
\end{figure}

\autoref{fig:stats-Eu-series} displays the evolution in energy and polarization across the FE phase transitions for the Eu series. We find a chemical trend in each of the three properties, with polarization and energy well depth increasing while band gap decreases as the B site moves down the periodic table of group 5 from P $\xrightarrow[]{}$ As $\xrightarrow[]{}$ Sb $\xrightarrow[]{}$ Bi. 
The key data are also summarized in \autoref{tab:stats}. 
As previously observed in another member of the 
A4B2C\blank{}tI14\blank{}139\blank{}e\blank{}a\blank{}ce\blank{}\blank{}+\blank{}-\blank{}- 
polaritype, \ce{Ba4Sb2O} \cite{markov2021}, we found the polar mode in both Eu and Yb families to be sensitive to epitaxial strain.
In the Yb series, compressive epitaxial strain of 1.5\% drives all four members to exhibit unstable polar modes, making them candidates for strain-induced ferroelectricity.
In both families, the critical amount of strain required to induce the ferroelectric state correlates with the B anion size, with smaller B anions requiring higher strain values.
However, once the ferroelectric phase is stable, further strain linearly  affects the ferroelectric mode frequency (\autoref{fig:fe-modes-strain}), with small B anions exhibiting steeper slopes.
In summary, compressive epitaxial strain destabilizes the ferroelectric mode, with the \ce{Yb4Pn2O} family featuring thresholds for the onset of ferroelectricity ranging between 0.5\% and -1.5\% strain.

\begin{figure*}
    \centering
    \includegraphics[width=0.85\linewidth]{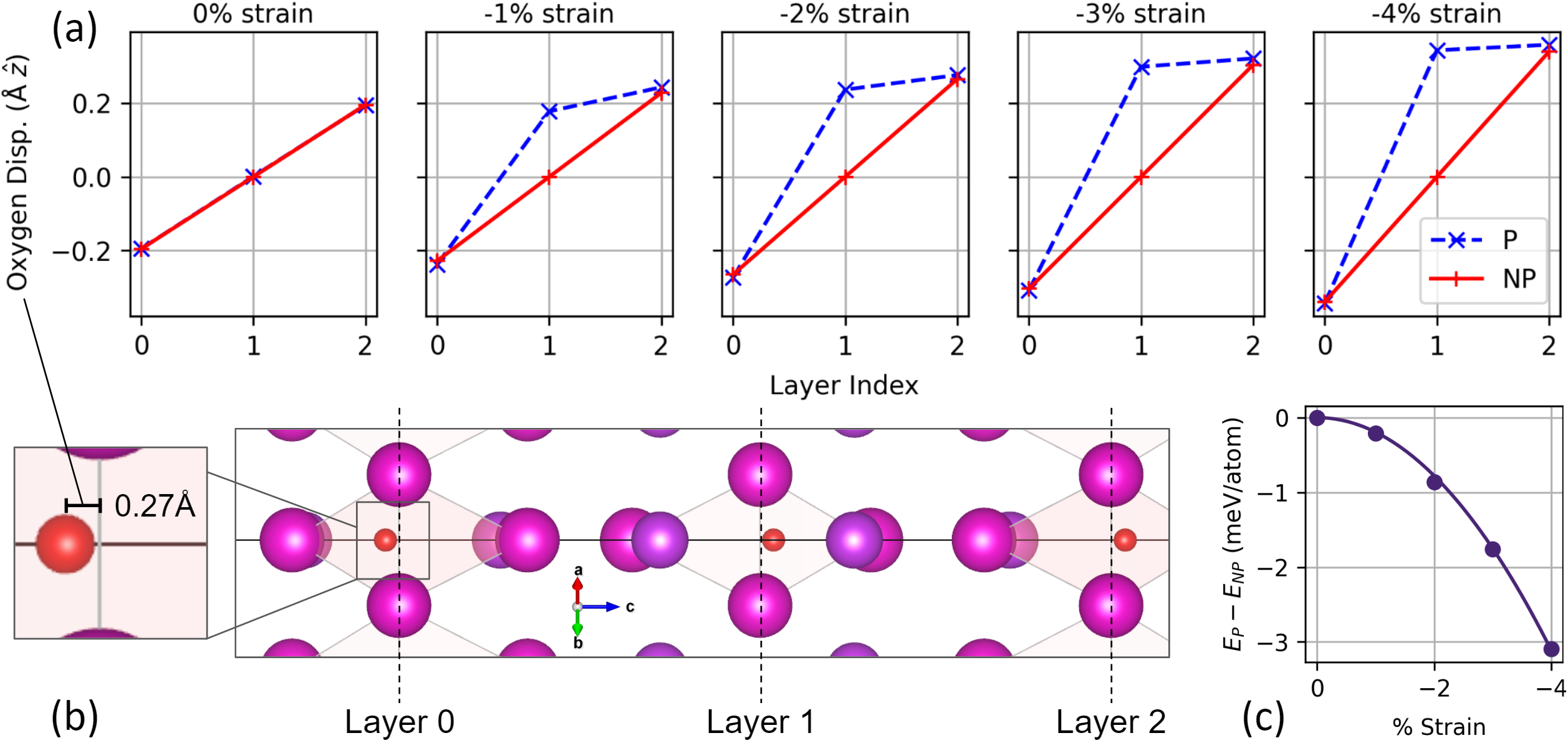}
    \caption{(a) Plots of the oxygen displacements in the strained \ce{Eu4Bi2O} films. (b) Illustrations of the oxygen displacement  measurement in the -4\% strain case. Each point represents the displacement (along the $c$ axis) of an oxygen atom from the basal plane of the \ce{OEu6} octahedron. (c) Plot of the relative energies of the polar (P, ground state) and nonpolar (NP, reference) \ce{Eu4Bi2O} structures as a function of epitaxial strain.}
    \label{fig:o-disps}
\end{figure*}

In the Eu series, compressive epitaxial strain further destabilizes the ferroelectric mode, as shown by the increasing imaginary frequencies as the strain goes from 0 to -4\% in \autoref{fig:fe-modes-strain}.
In \autoref{fig:phon-strain}, we show that the destabilizing effect of strain applies not only to the TO mode (\autoref{fig:fe-modes-strain}), but also to the LO mode.
In the top row of \autoref{fig:fe-modes-strain}, we see that increasing strain lowers both the TO and LO modes, with the LO mode also becoming unstable at 2\% strain.
An unstable LO mode is an indicator of hyperferroelectricity, or persistent out-of-plane polarization even in thin film geometries.

To further support the discovery of strain-activated hyperferroelectricity in \ce{Eu4Bi2O}, we calculate the structure and energy in a thin film geometry.
\autoref{fig:o-disps}a shows the position of the 3 oxygen atoms in the \ce{Eu4Bi2O} film -- consisting of 3 perovskite layers as shown in \autoref{fig:o-disps}b.
With no strain, the film exhibits an antipolar arrangement with no net polarization.
At 1\% or more compressive strain, the surface layers remain antipolar but the inner oxygen displaces along $c$ to generate a net polarization.
This antipolar surface phenomenon, which has been observed in other ferroelectric thin films, arises from the out-of-plane polarization causing a buildup of bound charge at the surfaces \cite{hwang2012,huyan2021,ohtomo2004}. 
This two-dimensional electron gas (2DEG), creates a depolarizing field which in this case is counteracted by the antipolar displacement of the oxygen anions in  \ce{OEu6} octahedra in the surface layers. 
In \autoref{fig:o-disps}a this is shown by the opposing displacement values in layer indices 0 and 2.
In \ce{Eu4Bi2O}, a single perovskite layer on each surface is enough to enable a persistent polarization in the central layer (shown by the positive displacement values for the polar structures -1\% and below in \autoref{fig:o-disps}a. 
In practice, we cannot probe a thinner layer of \ce{Eu4Bi2O}, because the film needs at least two perovskite layers to capture the alternating rock salt-perovskite layering characteristic of the Ruddlesden-Popper structure. In addition, the film must also allow for a centrosymmetric phase, which requires an odd number of perovskite layers.
\autoref{fig:o-disps}c shows that, like the polarization, the energy difference between the polar and nonpolar structures increases with strain.
In summary, \autoref{fig:o-disps} suggests that strain can be used to tune the extent of the oxygen displacement and energy well depth of the ferroelectric mode in a \ce{Eu4Bi2O} thin film.

We note that the  thresholds for strain-activated hyperferroelectricity is inconsistent between our two approaches: bulk phonons (\autoref{fig:phon-strain}) and thin film simulations (\autoref{fig:o-disps}).
In our film simulations, we also found that the \ce{Eu4Bi2O} film structure exhibits an elongated $c$ lattice parameter (out-of-plane) by 
1.5\% compared to the bulk cell. 
This results in a softening of the LO mode and thus the activation of the hyperferroelectric state at smaller strain values than predicted by the phonons.

Since hyperferroelectricity is native to thin films rather than bulk morphologies, we calculate a more accurate bulk-derived estimate of the hyperferroelectric boundary by fixing the $c$ lattice parameter of the bulk cell to that which is found by fully relaxing the thin film. 
As shown in \autoref{fig:phon-strain}, the strain required to induce hyperferroelectricity was predicted between -1\% and -2\% for the uncorrected bulk phonons (upper row) but between 0\% and -1\% for the corrected bulk phonons using the film-derived $c$ parameter (lower row).
The phonon spectra performed with this corrected cell yield hyperferroelectricity predictions consistent with  thin film calculations as confirmed by comparing the ``bulk $c$'' phonon spectra in \autoref{fig:phon-strain} to the relative energies in \autoref{fig:o-disps}c, both of which show hyperferroelectricity induced between 0\% and -1\% strain.
Combined, the unstable LO modes shown in \autoref{fig:phon-strain} and the persistent polarization in the 3-layer structure shown in \autoref{fig:o-disps}b constitute strong evidence of strain-activated hyperferroelectricity in \ce{Eu4Bi2O}.

\section{Conclusion}

In this work, we introduced the decoratype materials taxonomy, which generalizes the concept of anti-structures by classifying structural prototypes by their Wyckoff site-specific properties. This framework enables a more granular understanding of structure-property relationships and offers a powerful tool for materials discovery and design.
To explore this new taxonomy, we compiled a comprehensive data set by integrating data from three major open-source materials databases and classified their corresponding polaritypes. Our statistical analysis revealed that only $\sim$1\% of structure types exhibit a unique polaritype, indicating that most structures can accommodate multiple cation/anion arrangements. However, structure types with multiple observed polaritypes remain rare, comprising less than 2\% of the data set, with only $\sim$1\% exhibiting three or more polaritypes.

To demonstrate the practical utility of the decoratype framework, we conducted a case study focused on the discovery of (hyper)ferroelectric materials. We developed a lightweight, physics-informed active learning workflow that leverages decoratypes to efficiently navigate data-sparse regions of chemical space. We constructed a database of 606 high-throughput DFT calculations of $n=1$ anti-Ruddlesden-Popper structures. The workflow was benchmarked against a baseline using the CSP metric, an empirical metric measuring the likelihood of substitution, and was shown to be robust even with as few as 10 positive examples. It outperformed the baseline and demonstrated extrapolative capability across chemical groups. This approach identified six novel ferroelectric candidates, including three strain-activated ferroelectrics and three strain-activated hyperferroelectrics.
The framework opens new avenues for the design of advanced materials with tailored functionalities, such as switchable polarization and metal-insulator transitions.
    
\begin{acknowledgments}

We would like to thank Dr.\ David Hicks for answering our questions about AFLOW-XtalFinder and Dr.\ Kevin Garrity for enlightening discussion about the phenomenological model of hyperferroelectricity. 
This work was supported by the National Science Foundation (NSF) under award DGE-1842165.
This work was also supported by the SUPeRior Energy-efficient Materials and dEvices (SUPREME) Center SUPREME, one of seven centers in the JUMP 2.0 consortium, a Semiconductor Research Corporation (SRC) program sponsored by DARPA.

This work was jointly supported by two compute resources: 
(1) Bridges-2 at Pittsburgh Supercomputing Center through allocation mat230008p from the Advanced Cyberinfrastructure Coordination Ecosystem: Services \& Support (ACCESS) program, which is supported by National Science Foundation grants \#2138259, \#2138286, \#2138307, \#2137603, and \#2138296 \cite{brown2021}
and (2) the computational resources and staff contributions provided for the Quest high performance computing facility at Northwestern University which is jointly supported by the Office of the Provost, the Office for Research, and Northwestern University Information Technology.

\end{acknowledgments}

\section*{DATA AVAILABILITY}
The data that support the findings of this article are available in \cite{supp} and online at \cite{supp_dryad}.

\bibliography{references}


\end{document}